\documentclass[]{aastex631}

\usepackage{subfigure}
\usepackage[T1]{fontenc}
\usepackage{CJK}

\begin{document}
\begin{CJK*}{UTF8}{gbsn}

\title{Spectral analysis of the X-ray flares in the 2023 outburst of the new black binary transient Swift J1727.8--1613 observed with Insight-HXMT}

\correspondingauthor{Jin-Yuan Liao (廖进元)}
\email{liaojinyuan@ihep.ac.cn}

\correspondingauthor{Shuang-Nan Zhang (张双南)}
\email{zhangsn@ihep.ac.cn}

\author{Jia-Ying Cao (曹佳颖)}
\affiliation{Key Laboratory for Particle Astrophysics, Institute of High Energy Physics, Chinese Academy of Sciences, 19B Yuquan Road, Beijing 100049, China} 
\affiliation{University of Chinese Academy of Sciences, Chinese Academy of Sciences, Beijing 100049, China} 
\author[0000-0001-8277-6133]{Jin-Yuan Liao (廖进元)}
\affiliation{Key Laboratory for Particle Astrophysics, Institute of High Energy Physics, Chinese Academy of Sciences, 19B Yuquan Road, Beijing 100049, China} 
\author[0000-0001-5586-1017]{Shuang-Nan Zhang (张双南)}
\affiliation{Key Laboratory for Particle Astrophysics, Institute of High Energy Physics, Chinese Academy of Sciences, 19B Yuquan Road, Beijing 100049, China} 
\affiliation{University of Chinese Academy of Sciences, Chinese Academy of Sciences, Beijing 100049, China} 
\affiliation{Key Laboratory of Space Astronomy and Technology, National Astronomical Observatories, Chinese Academy of Sciences, Beijing 100101, China}
\author{Hua Feng}
\author{Jin-Lu Qu}
\author{Liang Zhang}
\author{He-Xin Liu}
\affiliation{Key Laboratory for Particle Astrophysics, Institute of High Energy Physics, Chinese Academy of Sciences, 19B Yuquan Road, Beijing 100049, China}
\author{Wei Yu}
\author{Qing-Chang Zhao}
\author{Jing-Qiang Peng}
\affiliation{Key Laboratory for Particle Astrophysics, Institute of High Energy Physics, Chinese Academy of Sciences, 19B Yuquan Road, Beijing 100049, China} 
\affiliation{University of Chinese Academy of Sciences, Chinese Academy of Sciences, Beijing 100049, China}  
\author{Ming-Yu Ge}
\author{Lian Tao}
\author{Yan-Jun Xu}
\author{Shu Zhang}
\affiliation{Key Laboratory for Particle Astrophysics, Institute of High Energy Physics, Chinese Academy of Sciences, 19B Yuquan Road, Beijing 100049, China} 
\author{Zi-Xu Yang}
\affiliation{School of Physics and Optoelectronic Engineering, Shandong Unversity of Technology, Zibo 255000, China}




\begin{abstract}

The new black hole transient Swift J1727.8--1613 exhibited a series of X-ray flares during its 2023 outburst extensively observed with \textit{Insight}-HXMT. We analyze the spectra of the flaring period using a series of models consisting of a multi-color disk and several different non-thermal components, and several consistent conclusions are obtained among these models. 
First, Swift J1727.8--1613 was in the transition process from the hard intermediate state (HIMS) to the very high state (VHS) during the first flaring period (MJD 60197--60204), and afterwards it exhibited typical VHS parameter characteristics, such as high temperature of the disk inner radius and a steep power-law spectrum with a photon index of 2.6. 
Second, the flares in the VHS are characterized by a rapid increase in the flux of accretion disk, accompanied by a simultaneous rapid expansion of the inner radius, which
could be apparent if the accretion disk hardening factor varies significantly.
The strong power-law component during the VHS is likely produced by synchrotron self-Compton process in the relativistic jets, in agreement with the observed weak reflection component and lack of correlation with the disk component. 

\end{abstract}



\section{Introduction} \label{sec:intro}

X-ray binaries (XRBs) have been extensively studied over the past few decades. With the continuous expansion of the XRB sample, the understanding of XRBs continues to deepen in both observation and corresponding theoretical analysis. Black hole XRB (BHXRB) is a type of XRB composed of a black hole (BH) and a companion star. The companion star provides accretion material to the BH, resulting in various temporal and spectral observational features \citep{1973A&A....24..337S,1995ApJ...455..623C,2006ARA&A..44...49R}. Compared to neutron star XRB, the compact star of the BHXRB is a black hole without a hard surface, hence the radiation characteristics of the two are very different.

Apart from a few XRBs that have continuous X-ray emissions, most BHXRBs are transients. These transients are in a quiescent state (QS) for most of the time and are not visible in the X-ray band under the limitation of the current X-ray all-sky monitors, followed by outbursts every few months to years (even decades), during which the X-ray luminosity increases by several orders of magnitude \citep{2002ApJ...572..392B}. For a complete outburst, BHXRBs typically undergo a cycle from quiescence to the low hard state (LHS), then to the high soft state (HSS), and finally through the LHS return to quiescence, which is represented as a \textquotesingle q\textquotesingle-shaped track on the hardness-intensity diagram (HID) \citep{2001ApJS..132..377H}. In the LHS, the radiation of XRB is dominated by the non-thermal radiation with a power-law index of 1.7; while in the HSS, it is dominated by the multi-color blackbody disk, accompanied by a weak non-thermal component with a power-law index of 2.2. There are several different ways to define the states of BHXRBs, typically by combining spectral and temporal characteristics to highlight the distinctive features of a particular state compared to others \citep{2005Ap&SS.300..107H,2006csxs.book..157M}.

The evolution tracks on the HID vary greatly among different XRBs, and some XRBs even fail to complete the \textquotesingle q\textquotesingle -shaped evolution during the outburst, e.g., GX339--4, GRO J1655--40, and XTE J1550--564. Many XRBs exhibit clear band-limited noise and quasi-periodic oscillation (QPO) as they approach the end of the LHS, known as the hard intermediate state (HIMS). However, they do not subsequently enter the soft intermediate state (SIMS) or HSS, but instead enter the very high state (VHS) and exhibit a significant steep power-law (SPL) component in their spectra.

There is a strong correlation between QPOs and spectral states, with the presence of QPOs being one of the factors used to determine the VHS \citep{2006csxs.book..157M}. While many observations have provided various explanations for these phenomena, none have been fully confirmed. The hot inner flow model could be used to explain the origin of QPOs \citep{2007A&ARv..15....1D,2013ApJ...778..165V}, but some observations and theoretical analyses have suggested more complicated scenarios \citep{2019NewAR..8501524I,2022MNRAS.511..255N}. Recent magnetohydrodynamics (MHD) simulation has been able to reproduce these observational phenomena, demonstrating the importance of magnetic fields in the accretion process \citep{2011ApJ...736..107O,2013ApJ...777...11S}. Furthermore, QPOs are widely believed to originate from non-thermal components but show a strong correlation with blackbody flux, aiding in the understanding of accretion complexity and geometric properties. 

Swift J1727.8--1613 is a BHXRB candidate discovered in August 24th, 2023 \citep{2023ATel16208....1C}. 
It exhibits a very high flux and complex signatures of optical inflows and outflow attributed to the very close source distance of $D=3.4$~kpc \citep{2024A&A...682L...1M,2025A&A...693A.129M}.
The spectral analysis of NICER observations has given the equivalent hydrogen column $N_{\rm H}=0.41\times10^{22}$ $\rm atom~cm^{-2}$ and inclination $\theta=47.9^{\circ}$, and revealed an extreme Kerr BH existing in this XRB \citep{2023ATel16219....1D}.
Additionally, the temporal analysis with \textit{Insight}-HXMT data also suggested that the BH has a high spin \citep{2024MNRAS.529.4624Y}. The X-ray polarization observations indicate that the corona is located within the inner region of the accretion disk, and extends along the disk plane 
\citep{2023ApJ...958L..16V,2024ApJ...968...76I}. Based on the polarization measurements of the accretion disk in the soft and hard states, it is believed that the source has a moderate inclination angle of $30^{\circ}$--$50^{\circ}$ \citep{2024ApJ...966L..35S}.
In addition, there may be a jet-corona configuration that could account for the spectral results from different model trials \citep{2024ApJ...960L..17P}.

During the rising phase of the Swift J1727.8--1613 outburst, \cite{2024arXiv240603834L} conducted spectral evolution analysis using the disk reflection model, and obtained the result that in the LHS the inner radius of the disk is close to the innermost stable circular orbit (ISCO). In this paper, we study the \textit{Insight}-HXMT continuous observations of Swift J1727.8--1613 during the 2023 outburst. The results show that it did not make transition to the HSS or SIMS after the HIMS, but exhibited strong SPL component in its spectra. Its temporal characteristics are similar to those in the HIMS, both exhibiting relatively significant low-frequency QPOs \citep{2024MNRAS.529.4624Y,2024ApJ...968..106Z,2024ApJ...961L..42Z,2024ApJ...970L..33Y}. It is worth noting that a series of significant flares can be observed in the light curve in low-energy band (2--6 keV). A detailed spectral analysis of Swift 1727.8--1613 during the whole flaring period is conducted, and the states of the source during each flare period are obtained. In addition, the origins of flares in different states and the related physical mechanisms are also discussed.

The paper is organized as follows. In Sections \ref{sec:data}, we describe the data reduction. In Section \ref{sec:result} and \ref{sec:dis}, we present the main results of the data analysis and make the relevant discussion. Finally, the summary and conclusion are given in Section \ref{sec:summary}.

\section{Data Reduction} \label{sec:data}

\textit{Insight}-HXMT carries three scientific payloads with different energy band, i.e., 
Low Energy X-ray telescope (LE) for 1--15~keV, Medium Energy X-ray telescope (ME) for 5--30~keV and High Energy X-ray telescope (HE) for 20--250~keV, respectively \citep{2020SCPMA..6349505C,2020SCPMA..6349504C,2020SCPMA..6349503L}. The combination of three telescopes enables \textit{Insight}-HXMT to cover a broad X-ray energy range of 1-250 keV. Furthermore, \textit{Insight}-HXMT has a large effective area and does not suffer from detector pile-up effects. Thus it has advantages in long-term continuous observation of bright sources and the study of the evolution of XRB outbursts \citep{2020SCPMA..6349502Z}.

The \textit{Insight}-HXMT Data Analysis software {\tt HXMTDAS v2.05} is used to extract the data from LE, ME and HE, 
and the latest calibration database ({\tt CALDB2.07}) and background files are adopted to ensure the accuracy of data extraction \citep{2020JHEAp..27...14L,2020JHEAp..27...24L,2020JHEAp..27...44G,2020JHEAp..27...64L}.
According to the recommendation of \textit{Insight}-HXMT Data Reduction Guide, the basic parameters for the good time interval (GTI) selection are consistent for the three telescopes, e.g., $ELV>10^\circ$, $COR>8~{\rm GV}$, $ANG\_DIST<0.04^\circ$, $T\_SAA>300~{\rm s}$, $TN\_SAA>300~{\rm s}$.
For LE, there is an additional condition \textquotesingle $DYE\_ELV>30^\circ$\textquotesingle~that needs to be met to avoid the visible light contamination from the bright Earth.
Taking into account the advantageous detection energy ranges of the three telescopes, we choose the 2--8 keV for LE, the 8--28 keV for ME, and the 28--200 keV for HE.
The joint spectral analysis of the three telescopes is
performed with the classical tool {\tt XSPEC v12.12.1}.
The systematic error in spectral analysis is set to 0.5\%. 

\section{Results} \label{sec:result}

\begin{figure}[ht!]
\centering
\includegraphics[scale=0.3]{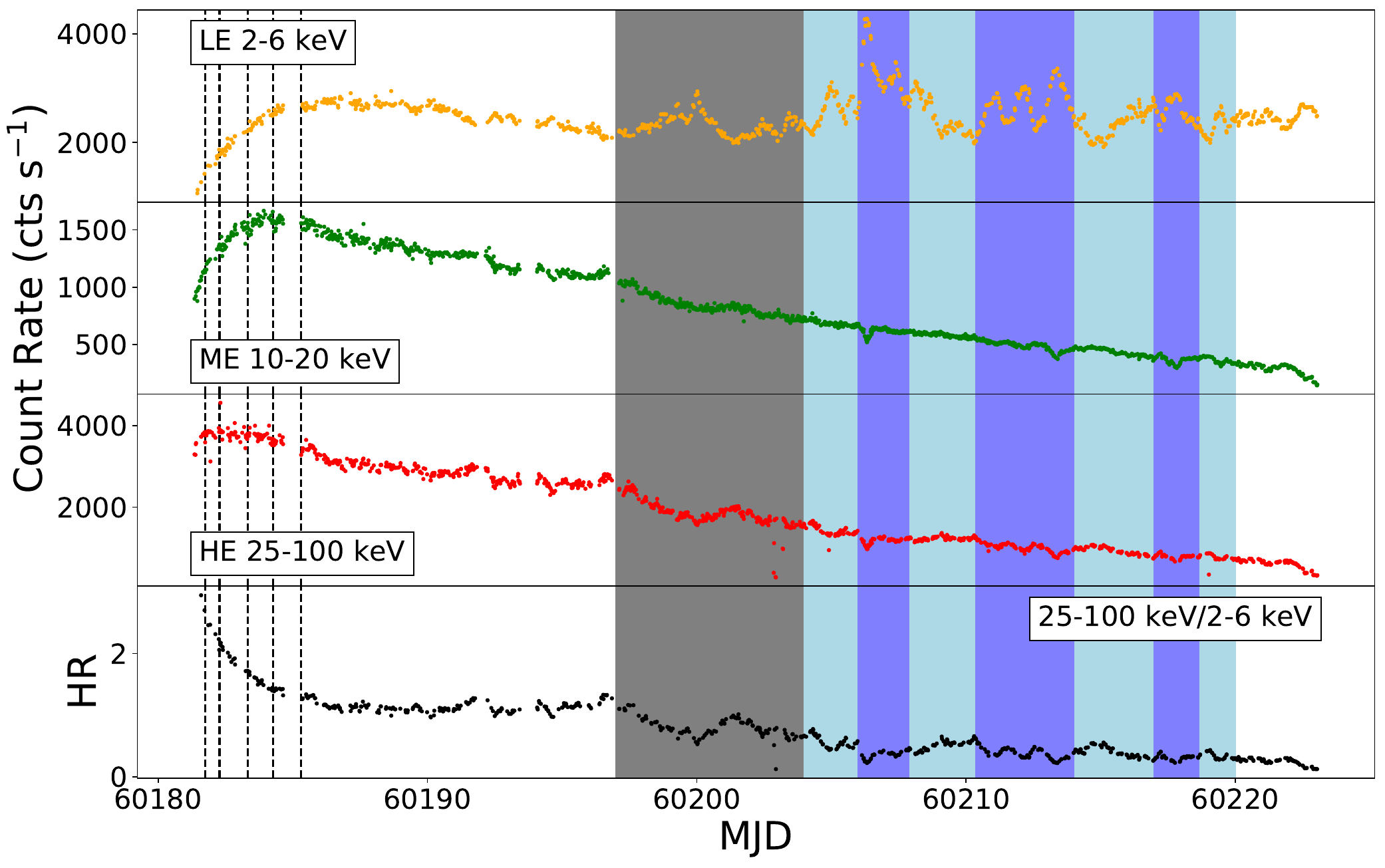}
\caption{
The top three panels are the light curves observed by \textit{Insight}-HXMT in 2--6~keV (orange), 10--20~keV (green) and 25--100~keV (red), respectively; while the bottom panel (black) is the evolution of the HR (25--100~keV/2--6~keV). 
The gray area represents the first flare period, the light blue area represents the weak flare period, and the dark blue area represents the strong flare period.
The black vertical dashed lines mark the five epochs analysed in \cite{2024arXiv240603834L}.
\label{fig:lcurve}}
\end{figure}

\begin{figure}[ht!]
\centering
\includegraphics[scale=0.3]{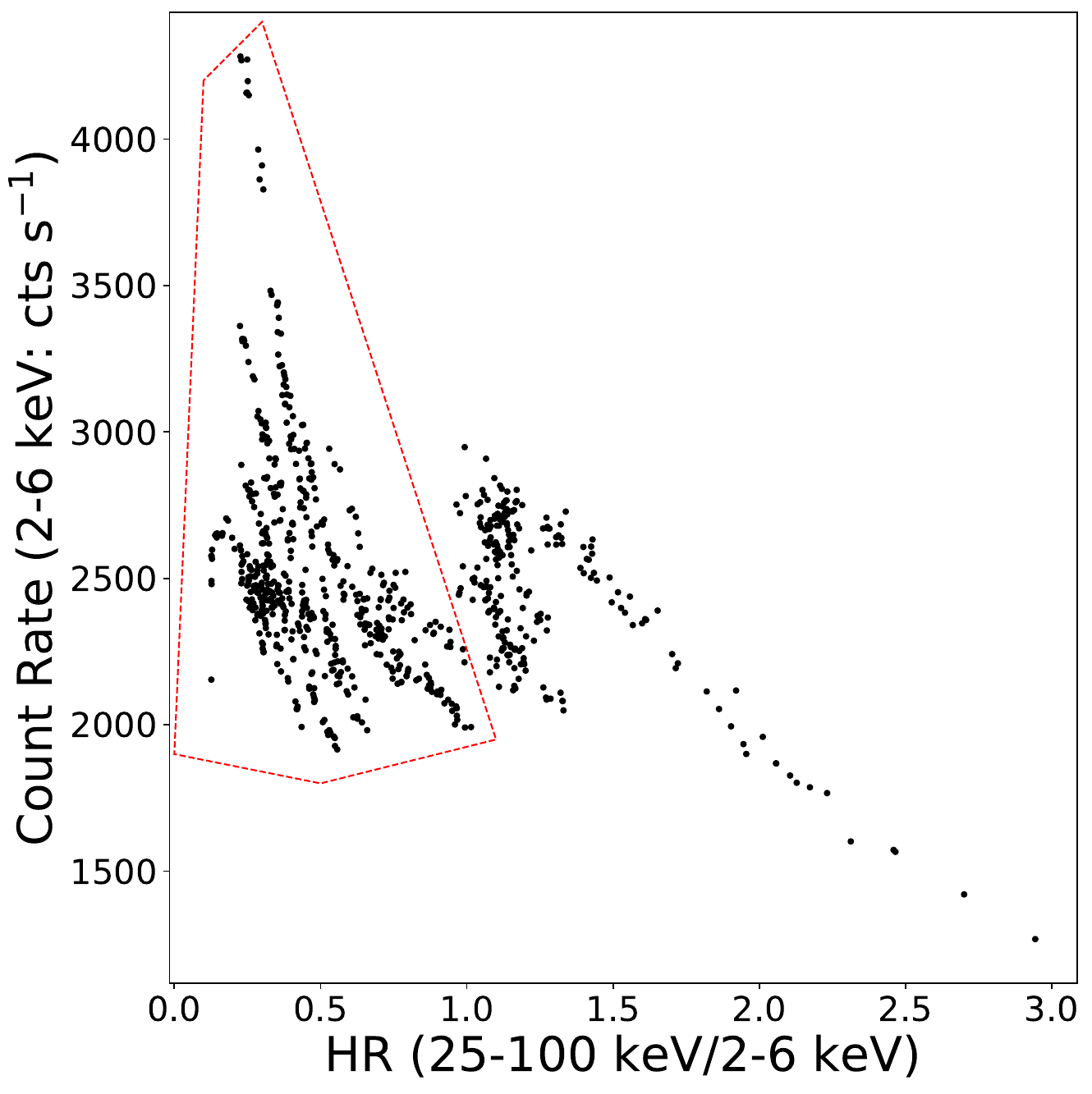}
\caption{
Hardness-intensity diagram constructed from \textit{Insight}-HXMT data. The red dashed box mark the selected time intervals studied in this work.
\label{fig:hid}}
\end{figure}

\textit{Insight}-HXMT monitored Swift J1727.8--1613 from August 25th, 2023 (MJD 60181) to October 4th, 2023 (MJD 60221) with a high cadence. The complete light curves of Swift J1727.8--1613 in three energy bands and the hardness ratio (HR: 25--100~keV/2--6~keV) are shown in Figure~\ref{fig:lcurve}.
There are five vertical dashed lines between MJD 60180 and 60190 indicating the low-hard state data studied by \cite{2024arXiv240603834L}.
After MJD 60185, the light curves of 2--6~keV have two types of behaviors, mainly divided into normal decay and flare periods.
The first significant flare in this light curve is roughly from MJD 60197 to 60204, followed by a series of flares continuing until MJD 60220.
However, the light curves of 10--20~keV and 25--100~keV continue to decrease in flux during the flare period, and some small dips appear at the peaks of the flares (around MJD 60200, MJD 60206, MJD 60213, and MJD 60218), which could be due to a temporary softening of the spectra caused by an increase of relative fraction of the disk component. For each flare period, the intensity of 2--6 keV increases and the hardness ratio decreases first and then go back.
We choose the data of \textit{Insight}-HXMT during the flare period (MJD 60197--60220) for spectral analysis, marked by the grey and blue area.
In Figure~\ref{fig:hid}, the HID exhibits the evolutionary trajectory during the continuous monitoring period of \textit{Insight}-HXMT, where a red dashed box marks the selected data.

\begin{figure}[ht!]
\centering
\subfigure{\includegraphics[scale=0.25]{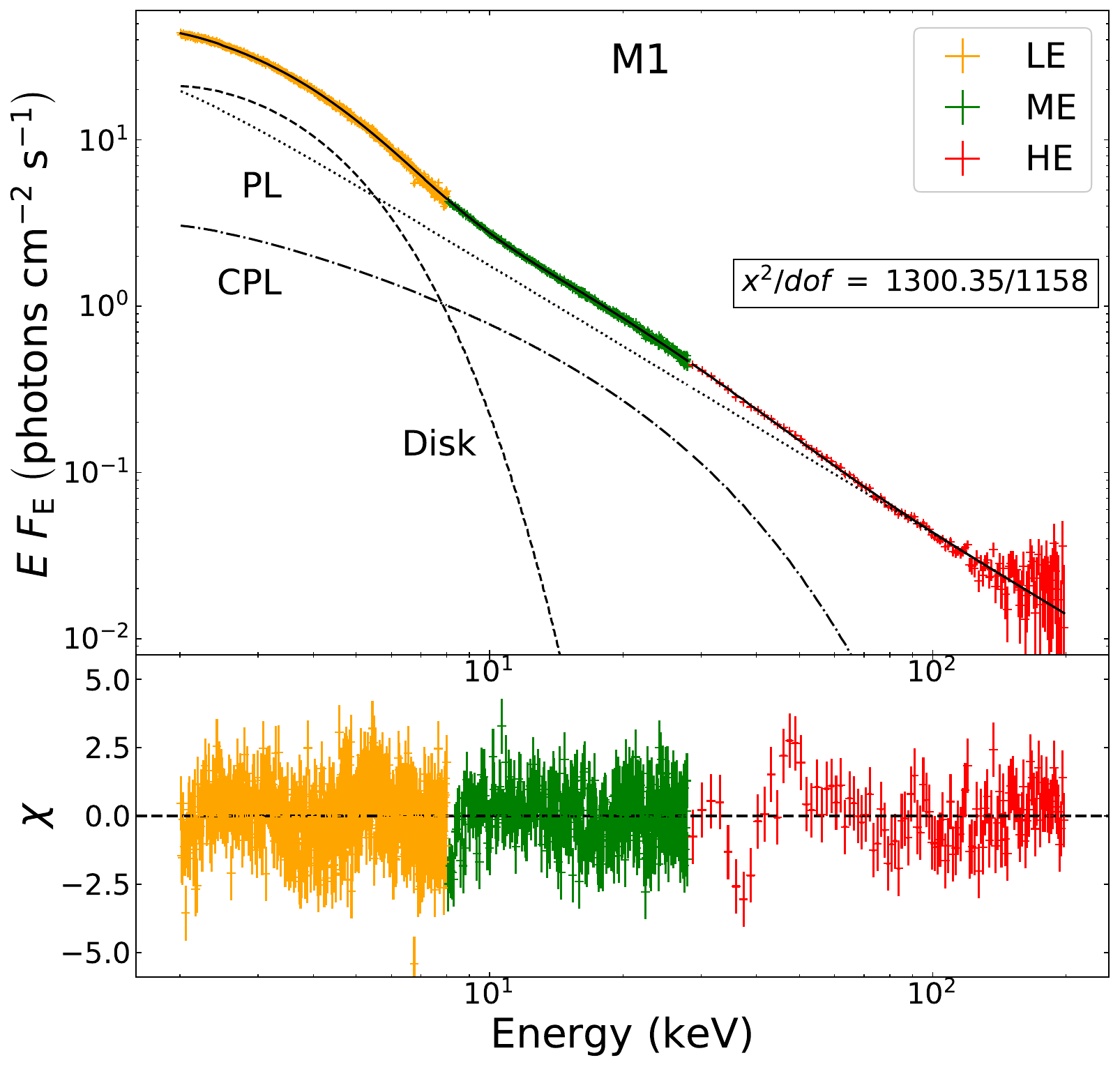}}
\hspace{.06in}
\subfigure{\includegraphics[scale=0.25]{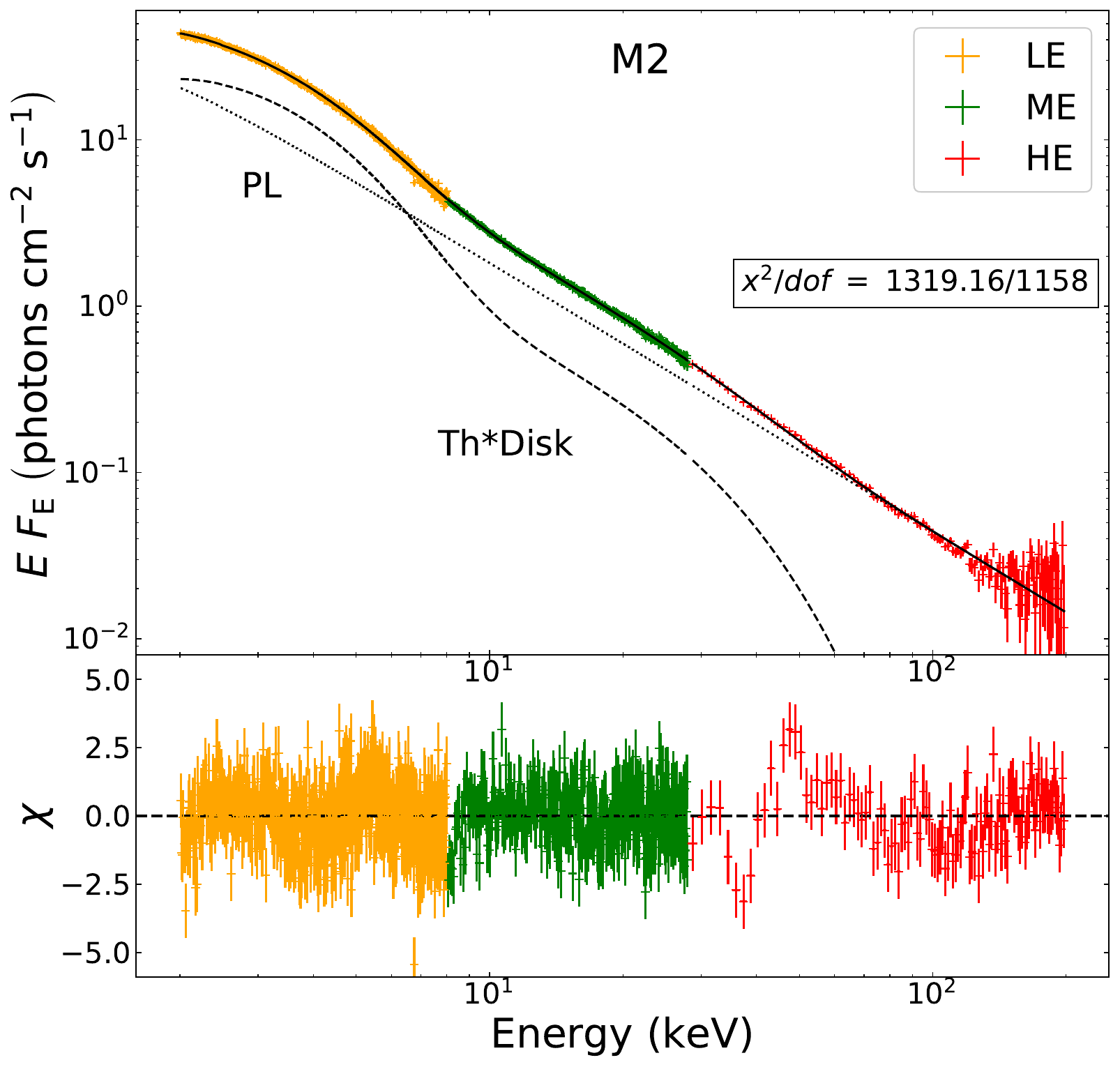}}
\hspace{.06in}
\subfigure{\includegraphics[scale=0.25]{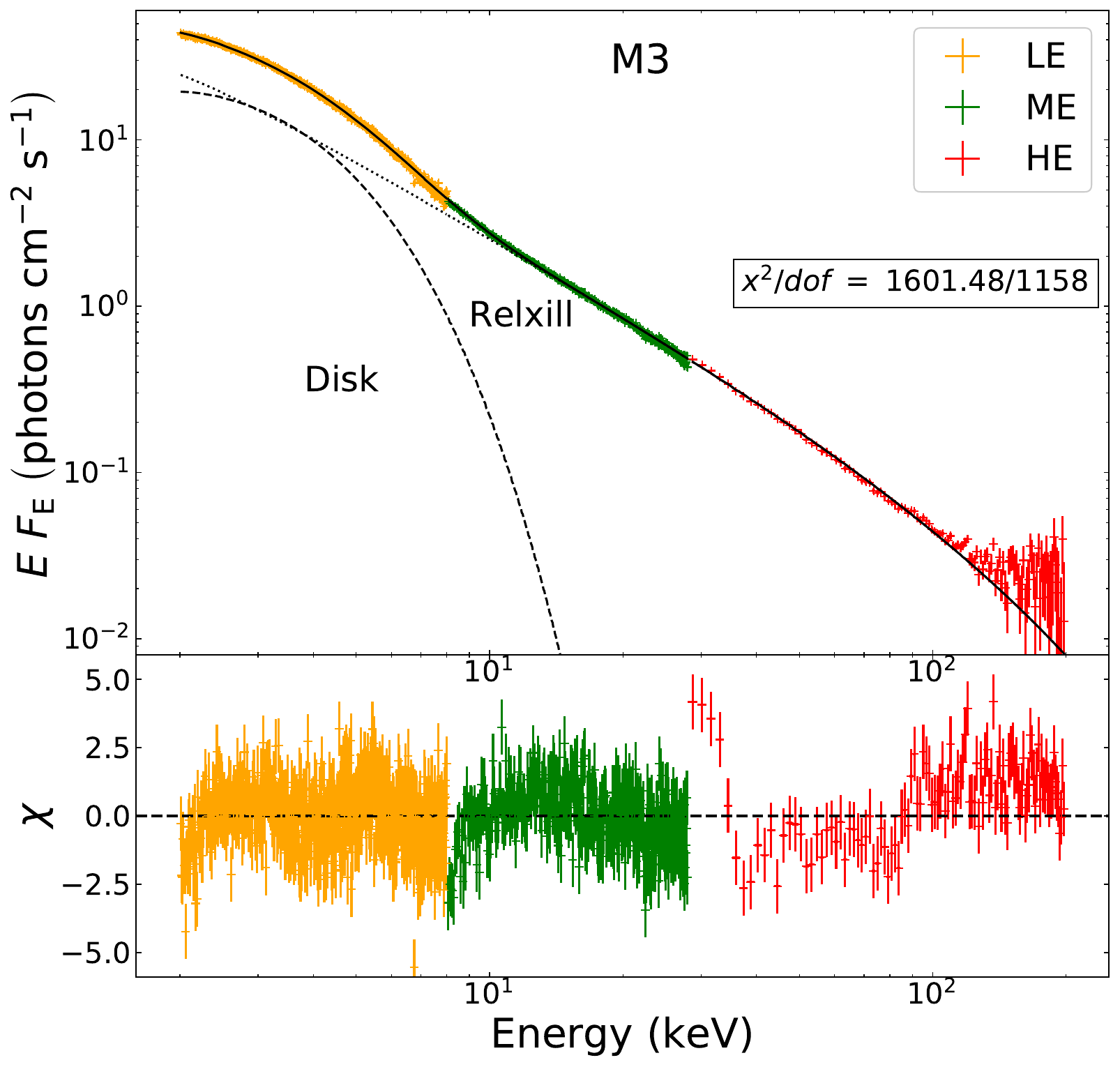}}
\hspace{.06in}
\subfigure{\includegraphics[scale=0.25]{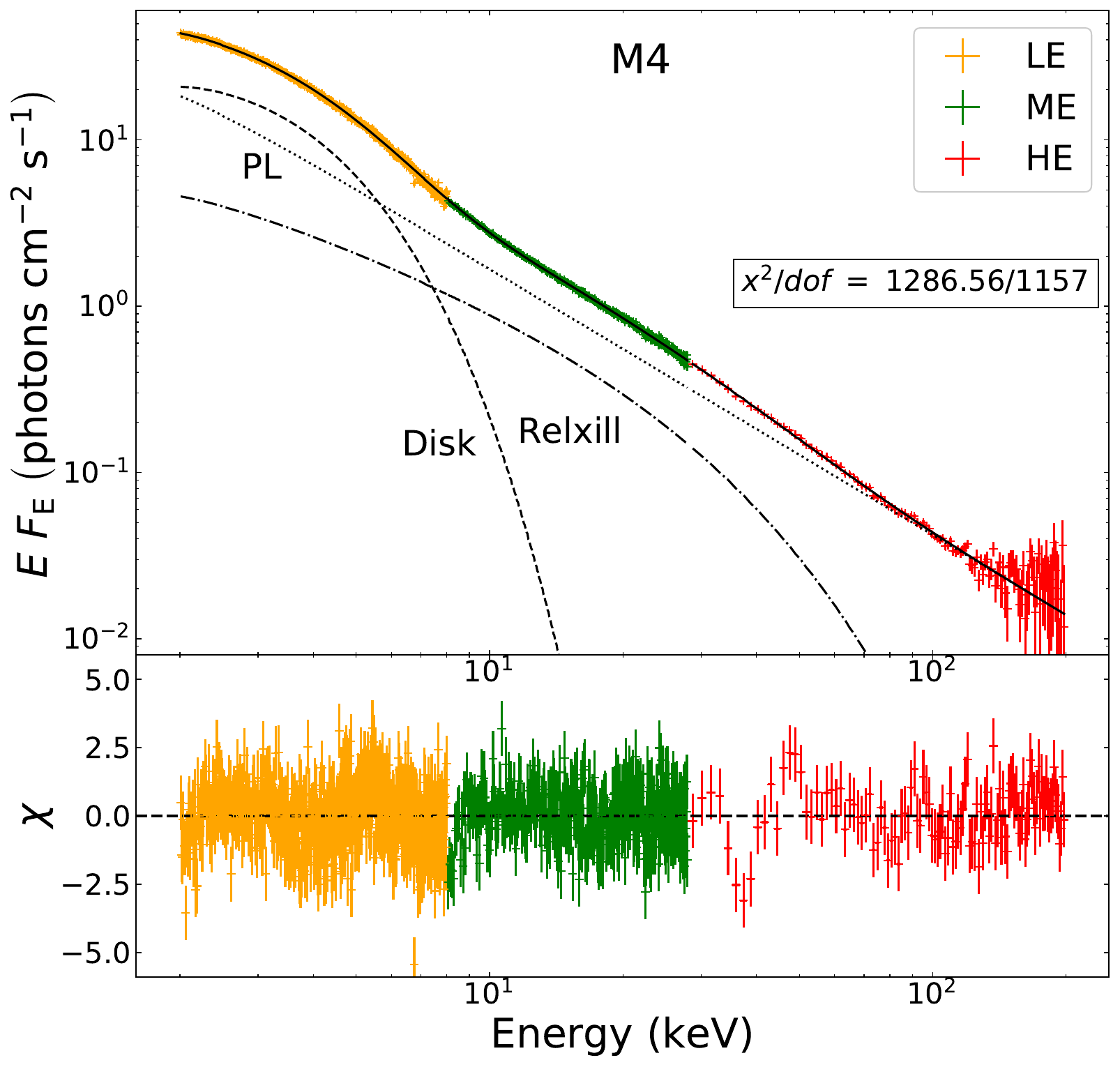}}
\caption{Examples of the observation P061433802003 fitted with models M1--M4.
The orange, green, red color represent the LE, ME and HE data, respectively.
\label{fig:spercum}}
\end{figure}

\subsection{Overall Spectral Evolution and State Transition} \label{sec:st}

Previous studies have shown that the X-ray energy spectra of BHXRBs exhibit both thermal and non-thermal components, both of which have different spectral properties in various states \citep{2006csxs.book..157M}.
In the spectral analysis of Swift J1727.8--1613, the thermal component is identified as a multi-color blackbody accretion disk, while the non-thermal component remains unidentified.
It is worth noting that the majority of the data cannot be well-fitted when only one non-thermal component is considered \citep{2024ApJ...960L..17P,2024arXiv240603834L}, which sets it apart from most of the BHXRBs.
We perform spectral analysis of 153 observations of \textit{Insight}-HXMT during MJD 60197 to 60220 with four spectral models, which all consist of a multi-color disk component \citep{1984PASJ...36..741M,1986ApJ...308..635M} and at least two non-thermal components. 
Additionally, each model includes a constant parameter for adjusting the response matrix of each detector, along with the Tuebingen-Boulder ISM absorption model \citep{2000ApJ...542..914W} utilized for interstellar absorption.
The composition of these four models, as well as the parameter description and settings for each component, can be found in Tables~\ref{tab:model} and~\ref{tab:parlist}.

\begin{deluxetable}{cc}
\tablecaption{Parameters of each model component\label{tab:model}}
\tablehead{\colhead{Component} & \colhead{Parameter} } 
\startdata
constant   &  An energy-independent multiplicative factor \\
\hline
tbabs      & $N_{\rm H}$: equivalent hydrogen column($\rm atom~cm^{-2}$) \\
\hline
diskbb     & \begin{tabular}[c]{@{}c@{}}$T_{\rm in}$: temperature at inner disk radius (keV) \\ $N_{\rm D}$: normalization factor \end{tabular}  \\
\hline
powerlaw   & \begin{tabular}[c]{@{}c@{}}$\alpha_{\rm PL}$: photon index \\ $N_{\rm PL}$: power law normalization factor \end{tabular}  \\
\hline
cutoffpl   & \begin{tabular}[c]{@{}c@{}}$\alpha_{\rm CPL}$: photon index \\ $E_{\rm cut}$: e-folding energy of exponential rolloff (keV) \\ $N_{\rm CPL}$: cut-off power law normalization factor \end{tabular}  \\
\hline
thcomp     & \begin{tabular}[c]{@{}c@{}}$\tau$: the Thomson
optical depth \\ $kT_{\rm e}$: electron temperature (keV) \\ $f_{\rm cov}$:  the scattering fraction \end{tabular}  \\
\hline
relxill    & \begin{tabular}[c]{@{}c@{}}\textit{Index1}\&\textit{Index2}: the emissivity indices \\ $R_{\rm in}$: inner radius of the accretion disk in gravitational radii \\ $R_{\rm out}$: outer radius of the accretion disk in gravitational radii \\ \textit{a}: spin of black hole \\ \textit{Incl}: inclination towards the system \\ \textit{z}: redshift to the source \\ \textit{$log~\xi$}: ionization of the accretion disk \\ $A_{\rm fe}$: the iron abundance of the material in the accretion disk \\ $\alpha_{\rm rel}$: power law photon index of the incident spectrum \\ $E_{\rm rel}$: high energy cutoff of the primary spectrum \\ $f_{\rm ref}$: reflection fraction \end{tabular}  \\
\enddata
\tablecomments{Fixed $N_{\rm H}=0.4\times10^{22}$, $\rm Index1\& \rm Index2=3$, $R_{\rm in}=-1$, $R_{\rm out}=400$, $a=0.998$, $z=0$, $\rm log~\xi=3.4$, $\rm Incl=47.9^{\circ}$, $A_{\rm fe}=0.5$} 
\end{deluxetable}

\begin{deluxetable*}{cc}
\tablecaption{\makebox[0.9\linewidth]{Compositions of the four models in spectral analysis}\label{tab:parlist}}
\tablehead{\colhead{Model} & \colhead{Component} } 
\startdata 
M1    & constant$\times$tbabs$\times$(diskbb+powerlaw+cutoffpl)   \\
\hline
M2    & constant$\times$tbabs$\times$(thcomp$\times$diskbb+powerlaw)  \\                                             
\hline
M3    & constant$\times$tbabs$\times$(diskbb+relxill)       \\
\hline
M4    & constant$\times$tbabs$\times$(diskbb+relxill+powerlaw) \\ 
\enddata
\end{deluxetable*}

The spectral fitting for the observations with the highest count rates in 2--6~keV with four models are presented as instances in Figure~\ref{fig:spercum}. In model M1, the two non-thermal components are a power-law continuum (PL) and a cutoff power-law continuum (CPL), respectively.
A hump around 10-30 keV is fitted by the CPL component, while a hard tail above 100 keV is fitted by the PL component, as shown in Figure~\ref{fig:spercum}(a).
In model M2, a model {\tt thcomp} is utilized to replace the CPL component, which describes a thermal Comptonization continuum formed by thermal electrons emitted by a spherical source \citep{1996MNRAS.283..193Z,2020MNRAS.492.5234Z}.
It can be seen that M1 and M2 have similar fitting goodness, and model M2 can provide a reasonable physical explanation for the CPL component of the M1 model. 
In addition, other possible physical mechanisms are also worth considering.
A relativistic reflection model {\tt relxill} is considered in model M3 and model M4 to describe the reflection of the photons from a corona hitting a disk \citep{2014MNRAS.444L.100D}.
Obviously, it is worth noting that the M3 model provides a poor fitting goodness compared to M1 and M2. The high-energy hard tail is not well-fitted in M3. Therefore, an extra PL is added in model M4 to fit the high-energy hard tail, where the fitting goodness has been greatly improved. The components used are all shown in Table \ref{tab:model}. 
Unlike the temperature at the inner radius ($T_{\rm in}$), which can be directly obtained from the spectral fitting, the realistic radius $R_{\rm in}$ comes from a series of conversions and corrections of the fitting parameter $N_{\rm D}$.
First, $N_{\rm D}$ is converted to the \textquotesingle fitted\textquotesingle~inner disk radius ($r_{\rm in}$) according to the method in \cite{1984PASJ...36..741M} with the source distance $D=3.4$~kpc and inclination $\theta=47.9^{\circ}$, and then $r_{\rm in}$ is corrected to $R_{\rm in}$ using the method in \cite{1998PASJ...50..667K} with $f_{\rm col}=1.7$ and $\xi=\sqrt{\frac{3}{7}}\left(\frac{6}{7}\right)^3$.
In the investigation of spectral analysis using models M1--M4, most of the model parameters are set free to explore the approximate range and trends of the parameters, as well as the degeneracy among these parameters. 
The evolution of parameters for the four models are presented in Appendix.

During the first flare period (MJD 60197--60204), the four models show that the non-thermal component dominates the flux while the disk component is very weak.
In model M1, $T_{\rm in}^{\rm (M1)}$\footnote{The superscript M1 refers to the parameter obtained from model M1 fitting. In this paper, superscripts are used to distinguish the same parameter under different models.} rapidly increases from $\sim0.5$~keV to $\sim1$~keV near the peak of the flare, while $R_{\rm in}^{\rm (M1)}$ shows an anti-correlation with $T_{\rm in}^{\rm (M1)}$. 
After the first flare, $R_{\rm in}^{\rm (M1)}$ exhibits a stable slow evolution with a small range of values, implying that $R_{\rm in}^{\rm (M1)}$ may be close to the ISCO. Additionally, the smaller value of $R_{\rm in}^{\rm (M1)}$ near the peak of the first flare can be explained by a larger hardening factor $f_{\rm col}$.
The values of $\alpha_{\rm PL}^{\rm (M1)}$ in the range of 2.0--2.6 are positively correlated with the PL flux and reach a maximum near the peak of the flare. Meanwhile, $\alpha_{\rm CPL}^{\rm (M1)}$ changes very little, maintaining a mean value of 1.66. $E_{\rm cut}^{\rm (M1)}$ fluctuates around 25~keV, and it appears to be positively correlated with $\alpha_{\rm CPL}$, which may be a result of parameter degeneracy.
The results of model M2 are considered unreasonable because $T_{\rm in}^{\rm (M2)}\sim3$~keV is significantly higher than the typical temperature range at the inner radius of an accretion disk.
The behavior of the disk in model M3 is similar to these in M1, with only small differences in specific parameter ranges. 
The flux of CPL ($F_{\rm CPL}^{\rm (M3)}$) and reflection components ($F_{\rm ref}^{\rm (M3)}$) obtained by separating the incident spectrum and reflection component in the relxill model for calculation, shows significant dispersion. This dispersion is attributed to the degeneracy resulting from the similar shapes of the two spectral components.
The evolutionary trend of $\alpha_{\rm rel}^{\rm (M3)}$ is very similar to that of $\alpha_{\rm PL}^{\rm (M1)}$, but with a smaller range of values.
It can be seen that the fitting goodness for most observations fitted with model M3 is statistically poor, suggesting the presence of an additional non-thermal component. Therefore, we continue to examine model M4, which is obtained by adding a PL component to model M3 and the results show that the spectral indices $\alpha_{\rm PL}^{\rm (M4)}$ and $\alpha_{\rm CPL}^{\rm (M4)}$ are both very close to those of M1.

During the first flare period, the parameters of the disk and PL components changed rapidly. Subsequently, in the following flares (MJD 60204--60220), the changes were relatively stable, with disk temperatures reaching around 1~keV and photon indices of $\sim2.6$. 
The timing analysis during this period indicates the presence of a series of Type-C QPOs, with their frequency increasing over time from $\sim 1$ Hz to $\sim 8$ Hz and rms of less than 15\%. These findings all support that the source underwent a transition from the HIMS to the VHS at the end of the first flare.
It is worth noting that a poor fitting goodness is obtained during the first flare period, indicating a very complex process of state transition. This will not be further elaborated in this paper, and a detailed analysis of the first flare will be conducted in our future work. We will focus on the spectral analysis of subsequent flares in the VHS in Section~\ref{sec:vhs}. Due to the parameter degeneracy identified in the previous analysis, which may lead to unexpected fluctuations in the fitting results, some parameters will be frozen in the subsequent spectral fitting to alleviate this parameter degeneracy.

\subsection{Spectral Evolution in the Very High State  and the Properties of the Flares} \label{sec:vhs}

Through the previous analysis, we find that the source entered the VHS after the first flare. During this period (MJD 60204--60220), there are a total of 101 observations by \textit{Insight}-HXMT. We continue to analyze these observations using the four models shown in Table~\ref{tab:parlist}, and fix some parameters in order to alleviate the possible parameter degeneracy. 
Same as Section \ref{sec:st}, the evolution of parameters for the four models during the VHS are shown in the Appendix.
In model M1, the parameter $\alpha_{\rm CPL}^{(\rm M1)}$ varies very little during the VHS with a mean value of $\sim$1.66, so we fix it to alleviate the degeneracy between PL and CPL.
According to the analysis results of model M1, it can be seen that the flux of the non-thermal component shows a decreasing trend throughout the entire VHS period, while the flux of the disk component shows an increasing trend.
The spectrum is dominated by the PL component with the photon index $\alpha_{\rm PL}^{\rm M1}$ slowly increasing from 2.6 to 2.65, accounting for approximately 60\% of the total flux in the energy range of 2--200 keV. The high proportion of the power-law spectral component and a photon index (>2.5) indicate that the source is in the VHS \citep{2006ARA&A..44...49R}. Throughout the entire VHS period, $T_{\rm in}^{(\rm M1)}$ shows a gradual decreasing trend, while the disk radius $R_{\rm in}^{(\rm M1)}$ exhibits an increasing trend. For the CPL component, $E_{\rm cut}^{(\rm M1)}$ increases from 20~keV to 30~keV. 

During the flaring intervals (around MJD 60209 and 60215), $T_{\rm in}^{(\rm M1)}$ decreases significantly and reaches a trough. However, during the flaring periods, $T_{\rm in}^{(\rm M1)}$ changes very little, and $R_{\rm in}^{(\rm M1)}$ increases significantly, leading to a rapid increase in the disk bolometric $F_{\rm D}^{(\rm M1)}$. 
Therefore, the flares are mainly a result of the rapid increase in disk bolometric flux attributed to the disk instability, 
which could be interpreted by an increase of the emitting region along with a rapid increase in $R_{\rm in}$, while $T_{\rm in}$ changes minimally. While if the rapid increase of $R_{\rm in}$ is apparent, it could mean the hardening factor deceased during the flare periods. The detailed discussion is in Section \ref{subsec:rin}.
There is a clear anti-correlation between $E_{\rm cut}^{(\rm M1)}$ and $F_{\rm D}^{(\rm M1)}$, which can be due to the Comptonization cooling of hot electrons by soft photons from the accretion disk. Further discussion will be provided in Section~\ref{sec:dis}.

The Comptonized accretion disk (model M2) has been investigated. 
From the spectral analysis with M2 in Section \ref{sec:st}, a potential degeneracy between $\tau$ and $kT_{\rm e}$ is observed, leading to a broader range of their values and making it difficult to determine.
To alleviate the degeneracy, we fix the electron temperature $kT_{\rm e}^{\rm (M2)}$ to half of the $E_{\rm cut}^{\rm (M1)}$, following the correlation between $kT_{\rm e}$ and $E_{\rm cut}$ in \cite{2001ApJ...556..716P}.
However, it can be seen that parameter degeneracy still exists from the similar evolutionary trajectories of $kT_{\rm e}$ and $f_{\rm cov}$.
The results of model M2 are very similar to those of model M1.
For the behavior of the accretion disk, the most significant difference between models M2 and M1 is the significant decrease in $T_{\rm in}$ and increase in $R_{\rm in}$ during the flare intervals (MJD 60209--60211 and MJD 60214--60215). 
For the Comptonization component, $kT_{\rm e}^{\rm (M2)}$ generally exhibited an increasing trend, but a decrease during the flare periods, while $\tau^{\rm (M2)}$ exhibited a completely inverse evolution. Moreover, $f_{\rm cov}^{\rm (M2)}$ exhibits an overall decreasing trend and a significant decrease during the flare intervals.
In model M2, the flares are also the result of an increase in accretion disk bolometric flux. A portion of the non-thermal components in the soft X-ray energy band are contributed by the inverse Compton scattering process. However, the electron temperature is only a few tens of keV, which significantly differs from the thermal electrons causing the high-energy tail.

In model M3, the diskbb model is used to fit the thermal component, while the reflection model relxill is employed to examine non-thermal radiation, which includes a CPL component and the resulting reflection on the disk. It initially exhibited a poor fitting goodness, but showed an improvement trend as the source evolved. As a result, an additional power-law component was added (model M4), leading to a significant improvement in fitting goodness. Detailed fitting results of models M3 and M4 can be found in the appendix. It is worth noting that for our data, the relxill model is not sensitive to $R_{\rm in}$.
Therefore, we simply fixed $R_{\rm in}$ in relxill to -1. If the BH mass is known, it can also be tied to $R_{\rm in}$ in diskbb for formal consistency, but the BH mass is currently unknown. Therefore, in both models M3 and M4, $R_{\rm in}$ is constrained by the diskbb model and remains consistent between relxill and diskbb.
The index of the PL component is fixed at 2.6, a value derived from the spectral analysis with models M1 and M2. The high-energy tail without a cutoff can be well fitted with the additional PL component.
In the reflection model, the CPL component not only contributes to the reflected component generated by the irradiated accretion disk but also contributes to a portion of the low-energy non-thermal radiation.
The characteristics of the disk component are similar to those in model M1, indicating that flares originate from an increase in disk bolometric flux.
Compared to model M1, model M4 includes an additional pure reflection component. The $E_{\rm rel}^{\rm (M4)}$ is higher than $E_{\rm cut}^{\rm (M1)}$ and continuously increases from $\sim20$~keV to $\sim100$~keV; while $\alpha_{\rm rel}^{\rm (M4)}$ slowly increases from 1.7 to 2.4. $f_{\rm ref}^{\rm (M4)}$ is less than 0.5 most of the time. It is inversely related to the disk bolometric flux and significantly decreases during flare periods.
By conducting an F-test on all spectral fittings with both models M3 and M4,
it can be found that adding a PL component in model M4 significantly improves the fitting goodness for most observations, except for some during the period of MJD 60217--60220.
This is because during this period, $E_{\rm rel}^{\rm (M3)}$ is around 200~keV, at which point the CPL and PL components are indistinguishable in the energy range of \textit{Insight}-HXMT. This suggests that in the end phase of the VHS, the inverse Compton process with only single-temperature electrons is sufficient to explain the observed non-thermal radiation.

In model M4, the accretion disk radiation in the VHS (MJD 60204--60220) is described by a non-relativistic diskbb model. To achieve better consistency with the relativistic reflection model relxill, the relativistic accretion disk model kerrbb \citep{2005ApJS..157..335L} is examined as a replacement for diskbb in M4. As a meaningful supplement to M4, the new model is designated as model M4k. Based on dynamical measurement \citep{2025A&A...693A.129M}, the black hole mass is fixed at 10 $M_\odot$, which is also adopted in \cite{2024ApJ...966L..35S}. The distance, spin, and inclination are still set to the values used in models M1--M4. The parameter evolution can be found in the appendix. It can be seen that the evolution of the flux of each spectral component is similar to that in models M1--M4, with the PL component dominating and the disk component significantly increasing during the flare period. In addition, the overall disk accretion rate ($\dot{M}$) shows an upward trend, with a significant increase during the flare period. The hardening factor $f_{\rm col}$ generally exhibits a downward trend and decreases significantly during the flare period.
There is a clear anti-correlation between $\dot{M}$ and $f_{\rm col}$. The spectral index and cutoff energy also generally show an upward trend, similar to the conclusions of models M3 and M4.

Combining the analysis results of these models above, we can draw some consistent conclusions. First, the flares are the result of an increase in disk bolometric flux in the VHS. The specific behavior of the accretion disk is characterized by minimal changes in $T_{\rm in}$ while $R_{\rm in}$ significantly increases in models M1--M4. In model M4k, it is exhibited a rapid increase in $\dot{M}$ and a significant decease in $f_{\rm col}$. Second, the spectrum in the VHS includes a strong PL component with a photon index of $\alpha_{\rm PL}\sim2.6$.
Third, both the scattering and reflection models provide statistically acceptable fitting goodness, with the best fit goodness observed in model M4. However, based on the spectral analysis of the parameter evolution in M4, it is noted that the reflection fraction is too small and the iron line is too weak to be fitted, indicating very weak reflection. In this case, the M4 model exhibits a component composition similar to that of the M1 model. Besides, the acceptable fit goodness in M2 suggests that the Compton scattering process between the disk and the corona is also worth discussing.
We discussed several possible scenarios for non-thermal components and provided schematic diagrams of potential models in Section \ref{subsec:non-th}. Additionally, the addition of an extra PL component in M4 significantly improves the fitting goodness for the first half of the observations. However, the similar fitting goodness between models M3 and M4 during the period of MJD 60217--60220 suggests that there may have been a change in the composition of the non-thermal component at the end of the VHS.

\section{Discussion} \label{sec:dis}

We have performed a spectral analysis of the X-ray flares in Swift J1727.8--1613 with the observations of \textit{Insight}-HXMT from MJD 60197 to 60220. The spectra are dominated by a PL component with a slowly increasing photon index $\alpha_{\rm PL}$, while the disk component has a relatively low contribution and shows slowly decreasing $T_{\rm in}$ and slowly increasing $R_{\rm in}$. We find that the first flare resulted from an increase in the flux of the PL component. The evolution of various spectral parameters indicates that the source was undergoing a transition from the HIMS to the VHS. The fitting goodness of the four models is unsatisfactory, indicating that the source has a complex radiation process during the state transition. We will conduct a detailed study on this in future work.
In this section, we mainly focus on the spectral characteristics of flares in the VHS (MJD 60204--60220). As the models with different non-thermal components can provide acceptable fits, careful analysis is needed to determine the origin of the non-thermal radiation. However, the results of the accretion disk and power-law component are similar in these models.

\subsection{Origin of the Flares}

\begin{figure}[ht!]
\centering
\subfigure{\includegraphics[scale=0.25]{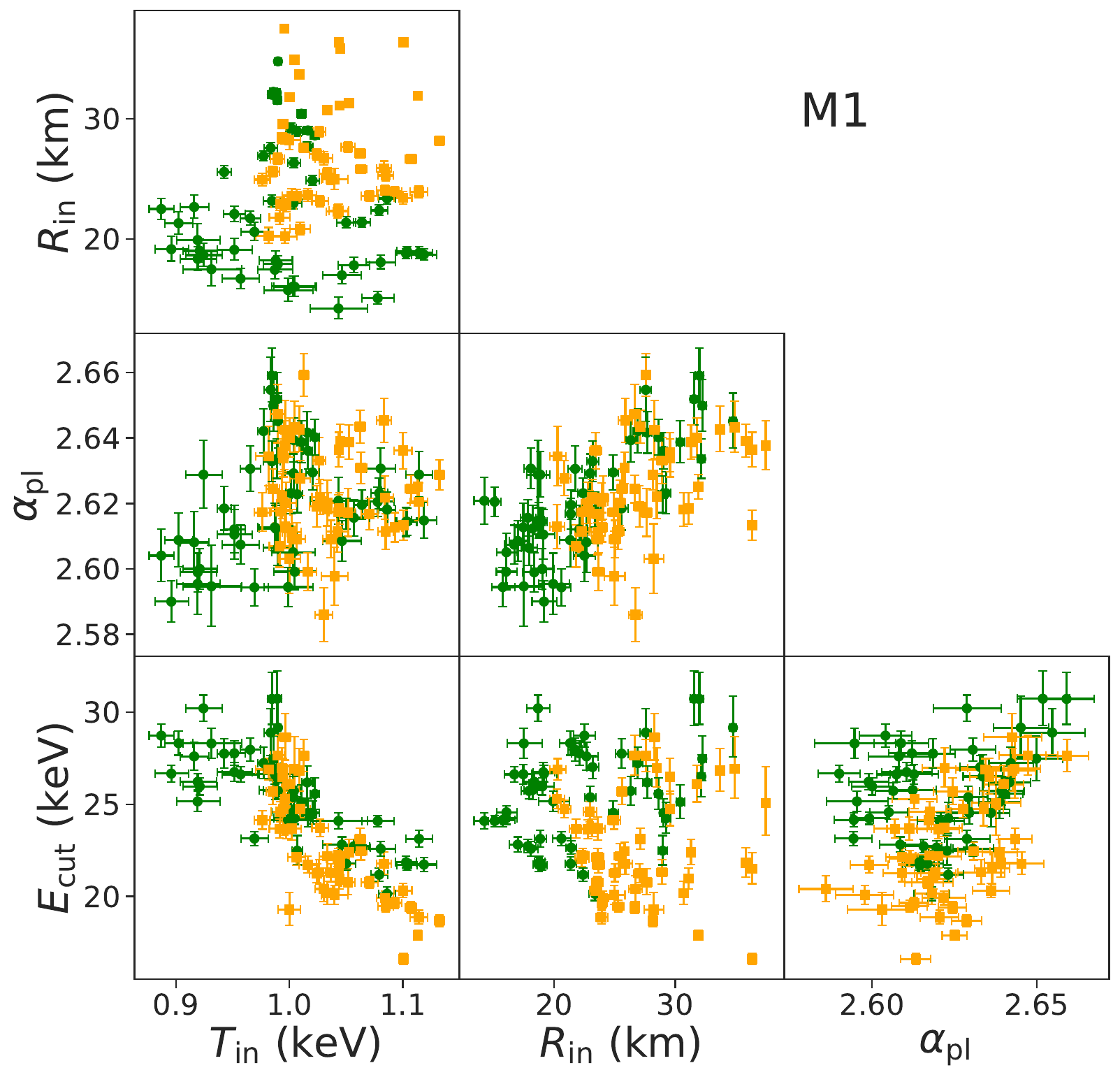}}
\hspace{.06in}
\subfigure{\includegraphics[scale=0.25]{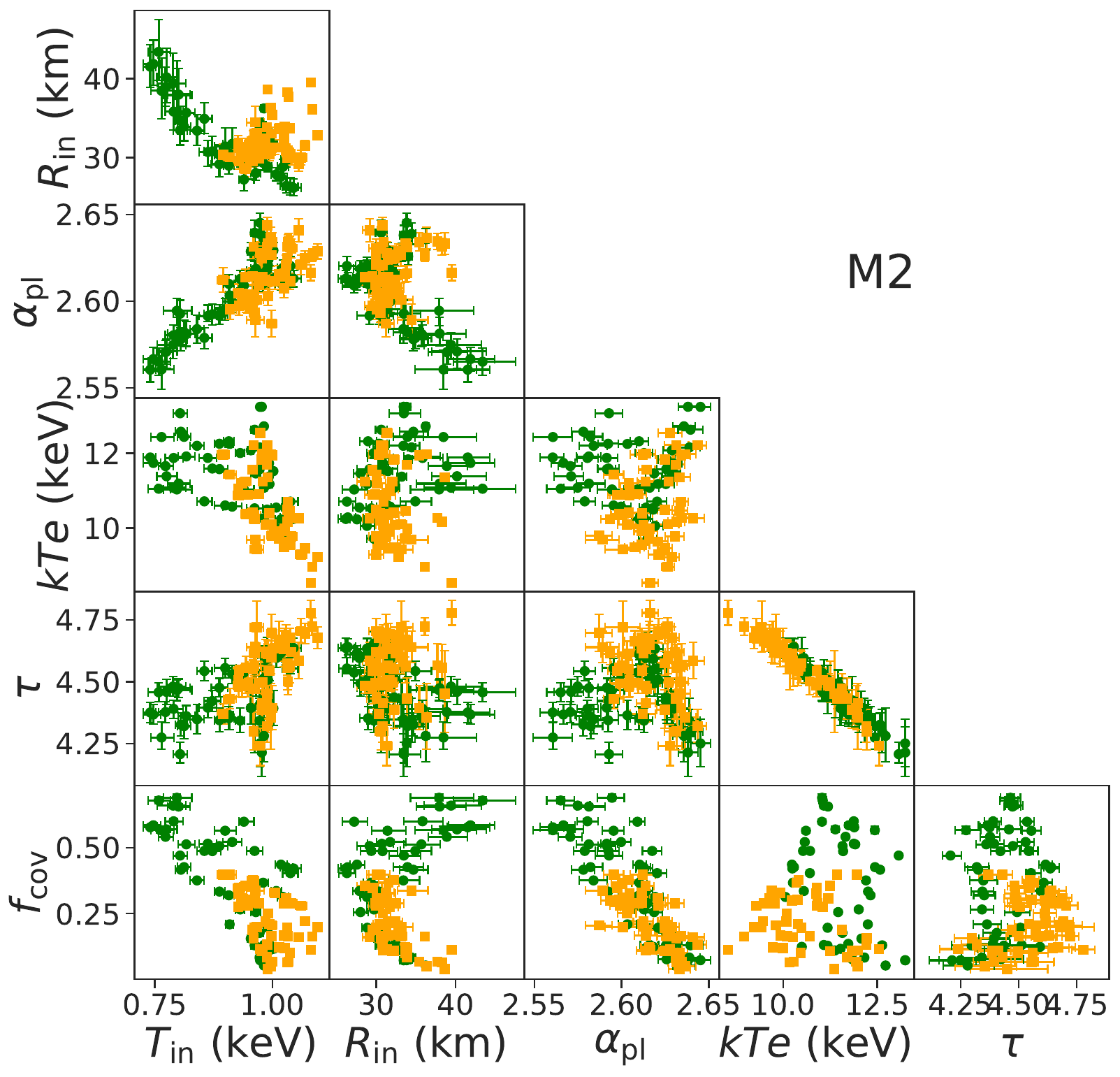}}
\hspace{.06in}
\subfigure{\includegraphics[scale=0.25]{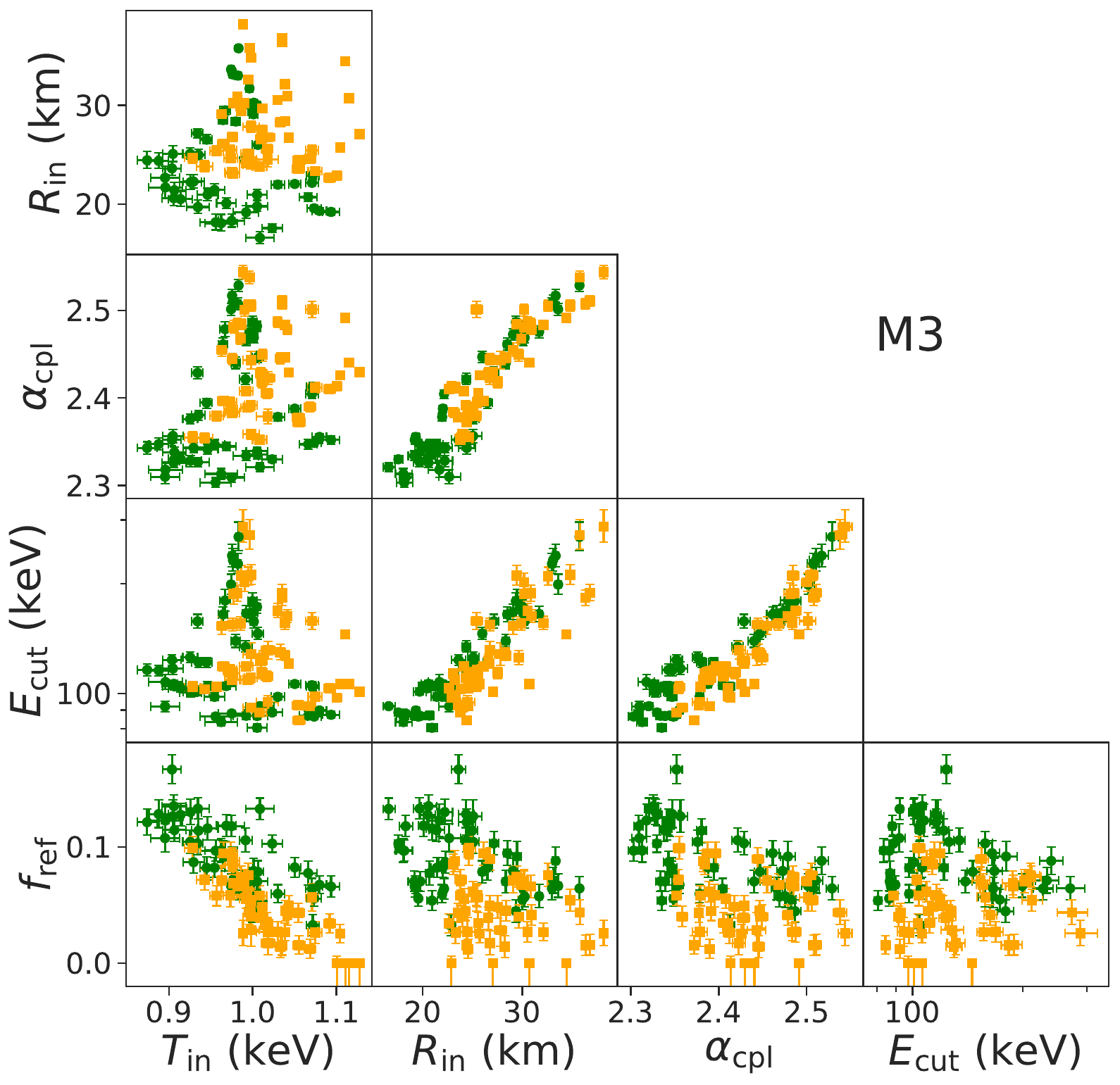}}
\hspace{.06in}
\subfigure{\includegraphics[scale=0.25]{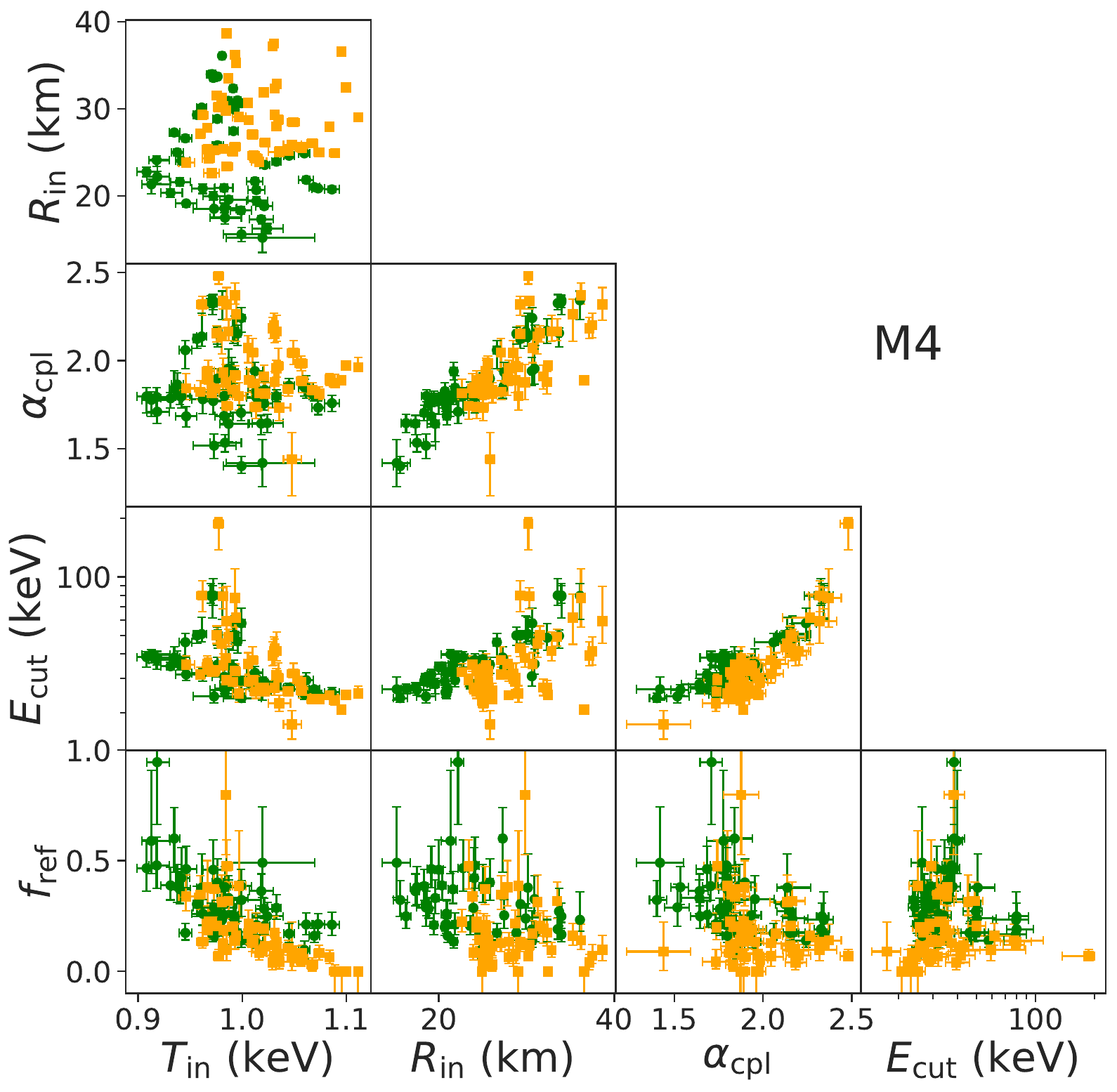}}
\caption{During the strong flare (orange color) periods and the weak flare (green color) periods, the correlation among the parameters with models M1--M4.
\label{fig:corr}}
\end{figure}

It can be seen that the variability is very pronounced following HIMS (Figure~\ref{fig:lcurve}). We simply defined the time intervals around significant peaks as the strong flare period, and the rest as the weak flare period.
To study the mechanism of flares, the relationship between various spectral parameters during strong and weak flare periods is examined (Figure~\ref{fig:corr}).
In model M1, the temperature and radius of the disk are generally higher compared to weak flare periods during the strong flare periods, and there is a significant anti-correlation between $T_{\rm in}$ and $E_{\rm cut}$. Comparing the correlation of parameters between strong and weak flare periods, it can be observed that there is no significant difference between the two. 
However, in model M2, significant differences are observed between strong and weak flare periods. During the strong flare periods, the trends of $T_{\rm in}$, $R_{\rm in}$, and $\alpha_{\rm pl}$ in M2 model are similar to those in M1 model. During the weak flare period, there is a strong correlation among these three parameters, possibly due to the parameter degeneracy. 
For the Comptonization component, the electron temperature remains at a high value during the weak flare periods, while during the strong flare periods, it shows an anti-correlation with the accretion disk temperature and radius. 
$f_{\rm cov}$ remains at a generally low level during the flare periods and is anti-correlated to the accretion disk temperature, radius, and the photon index of the PL component. The fraction of scattering component decreases and the parameters evolve more rapidly during the strong flare period compared to the weak flare period, indicating a rapidly changing corona.

In the bottom panel of Figure \ref{fig:corr}, the correlations of spectral parameters for M3 and M4 are plotted, with the photon index of the PL component in M4 fixed at 2.6. For model M3, there is a strong parameter degeneracy between the inner radius of the accretion disk and the photon index and cutoff energy of the reflection component. The cutoff energy can be as high as several hundred keV, and the reflection fraction is very small, indicating almost no reflection. 
It also shows a very weak reflection component during both strong and weak flare periods. This could be caused by the fact that the energy spectrum of the source is dominated by the PL component, while the component of the corona irradiating the disk is very weak. With the very weak reflection component in M4 model, M4 model exhibits a composition similar to that of M1. Additionally, a more detailed physical picture is provided in M2, consisting a Comptonized disk and a very strong hard tail.

\begin{figure}[ht!]
\centering
\subfigure{\includegraphics[scale=0.45]{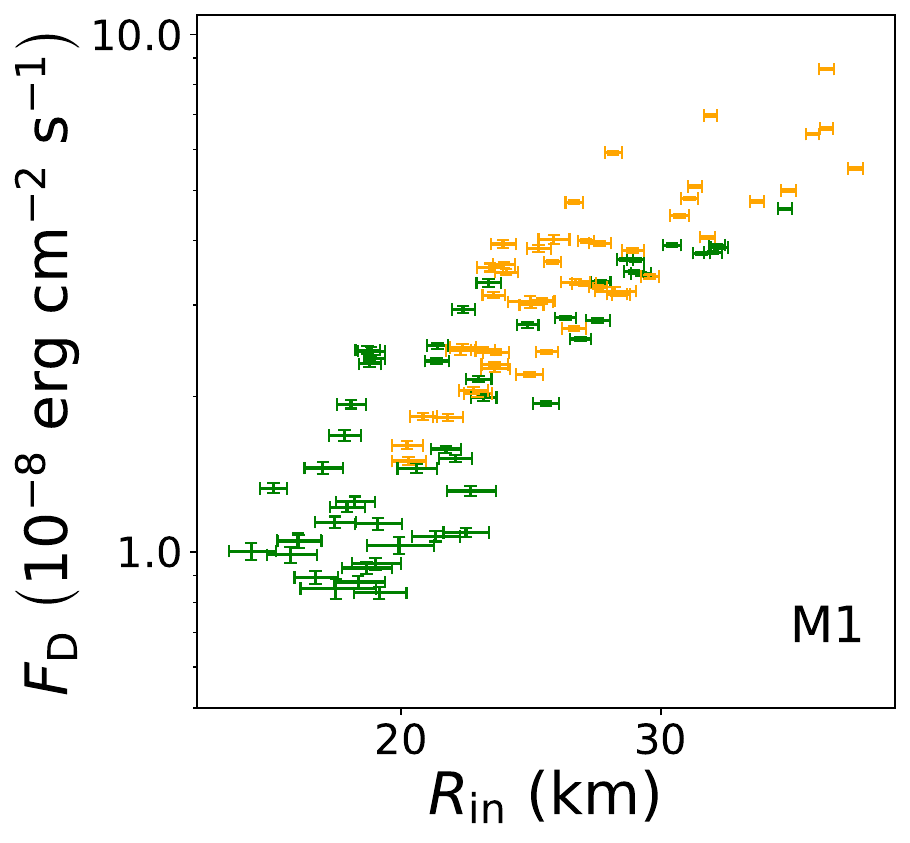}}
\hspace{.06in}
\subfigure{\includegraphics[scale=0.45]{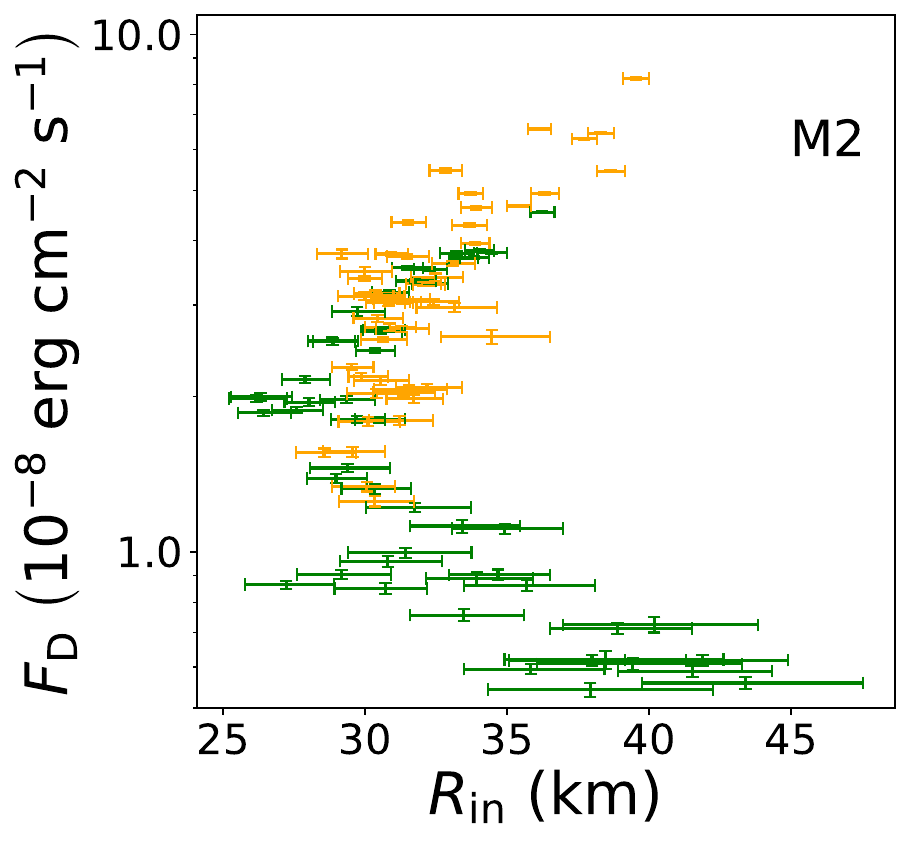}}
\hspace{.06in}
\subfigure{\includegraphics[scale=0.45]{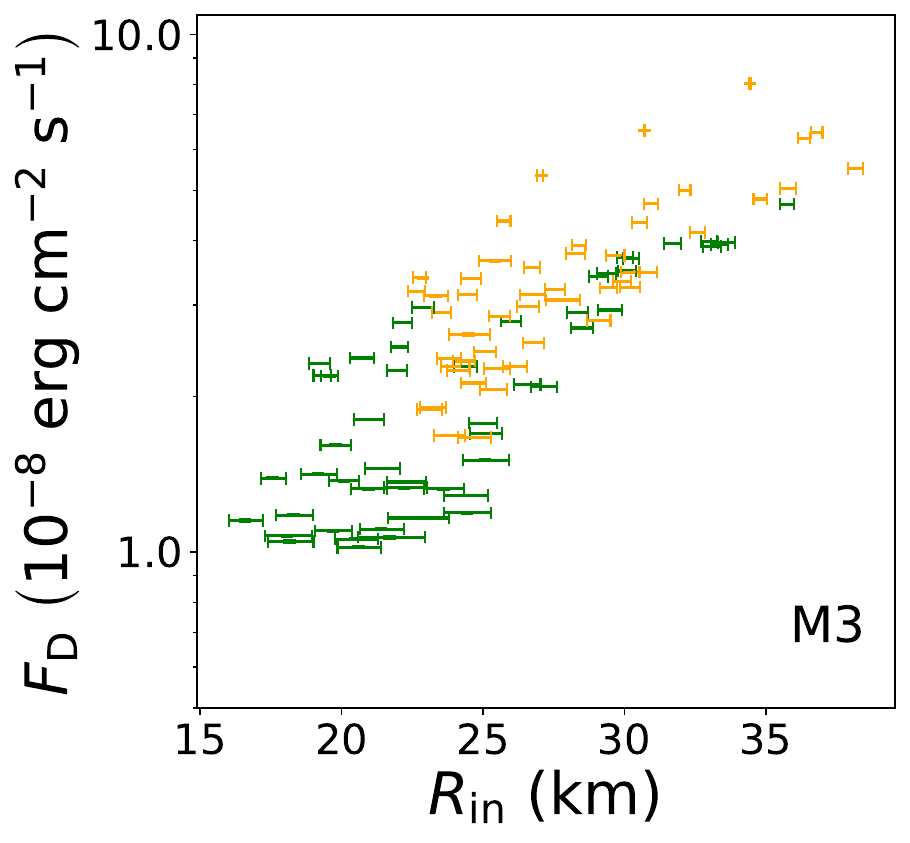}}
\hspace{.06in}
\subfigure{\includegraphics[scale=0.45]{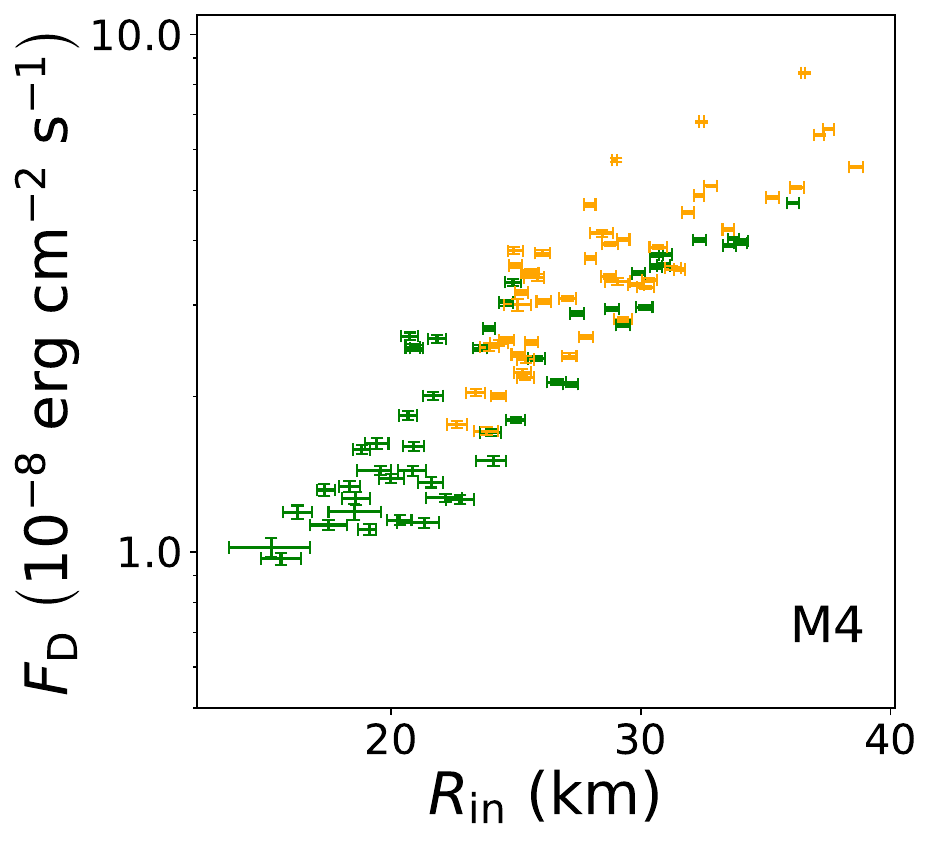}}
\caption{The relationship between the disk bolometric flux ($F_{\rm D}$) and the inner radius of multi-color disk ($R_{\rm in}$) in the spectral fitting of four models.
The data during the strong and weak flare periods are represented in orange and green color, respectively.
\label{fig:corrin}}
\end{figure}

\begin{figure}[ht!]
\centering
\subfigure{\includegraphics[scale=0.45]{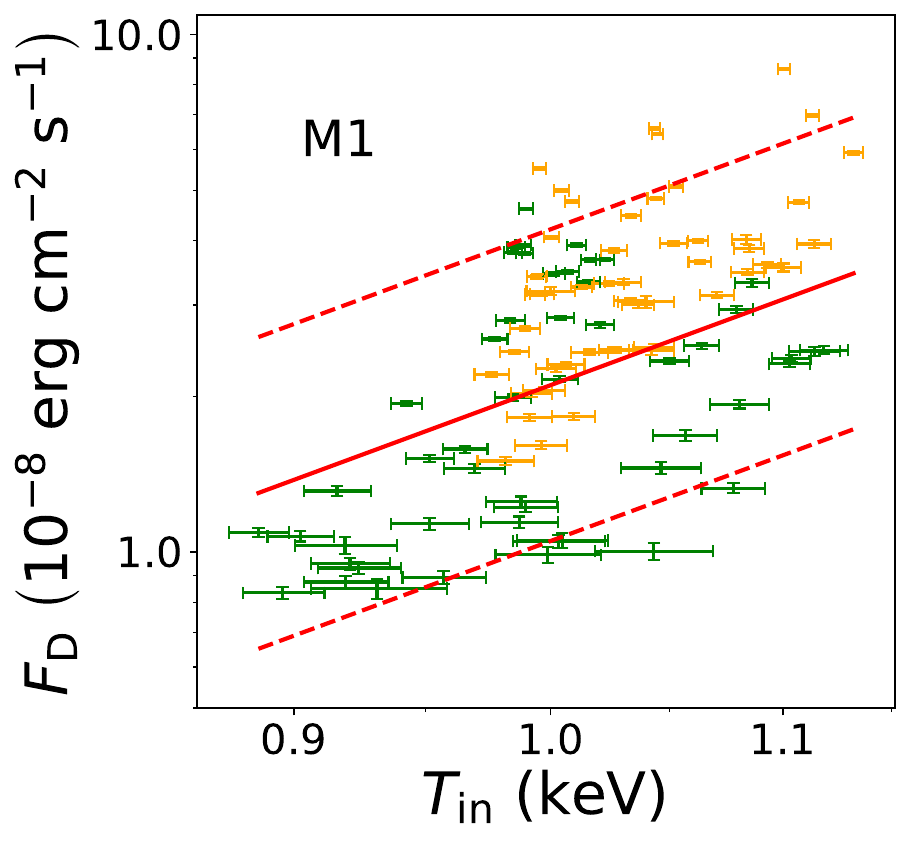}}
\hspace{.06in}
\subfigure{\includegraphics[scale=0.45]{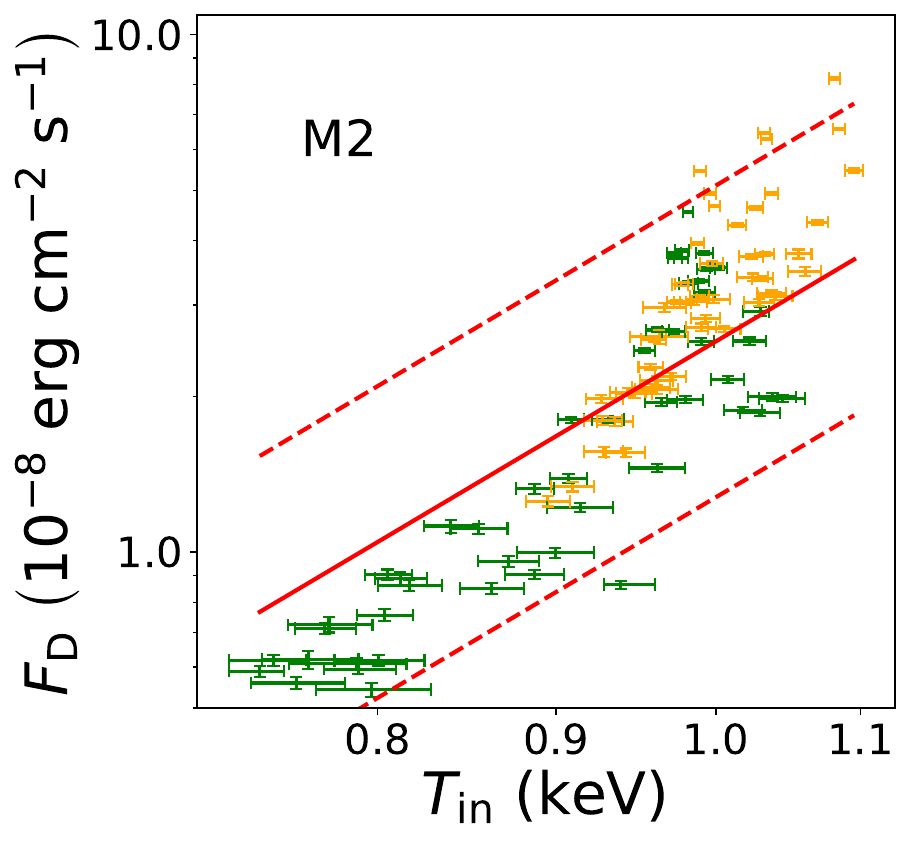}}
\hspace{.06in}
\subfigure{\includegraphics[scale=0.45]{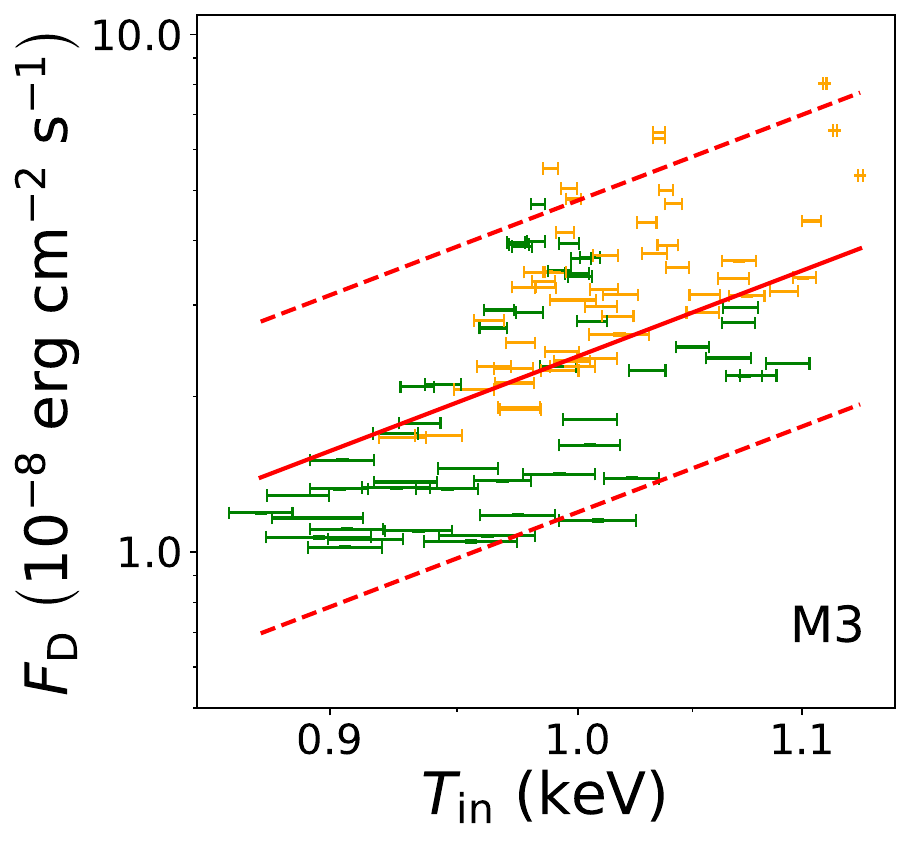}}
\hspace{.06in}
\subfigure{\includegraphics[scale=0.45]{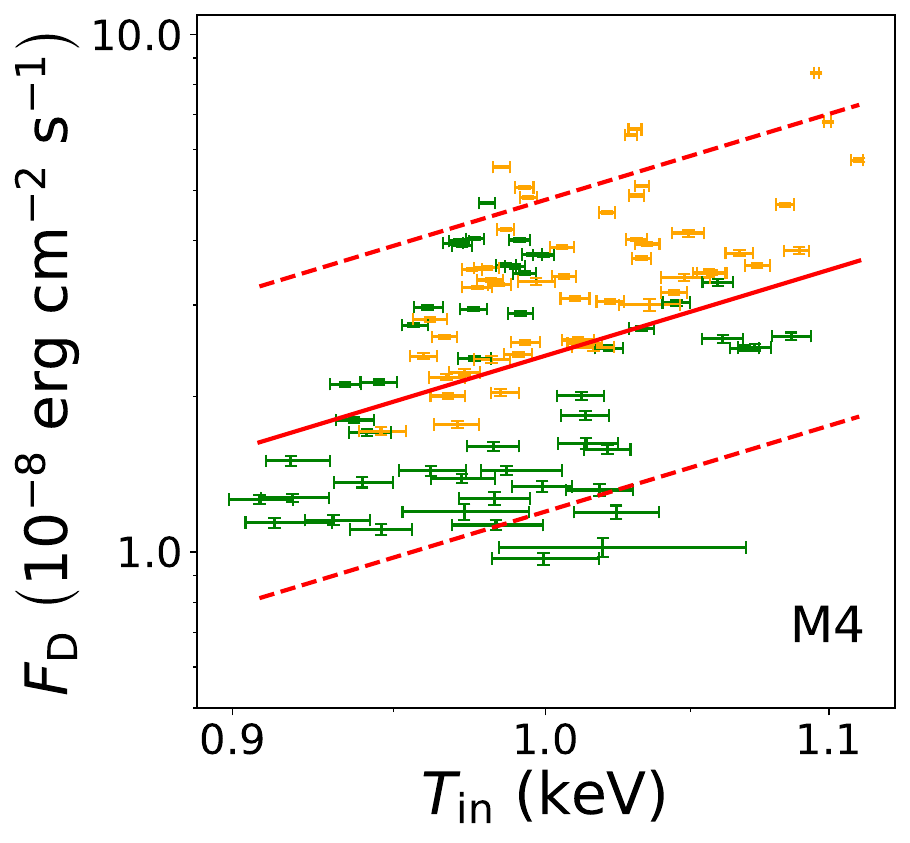}}
\caption{The relationship between the disk bolometric flux ($F_{\rm D}$) and the temperature of multi-color disk ($T_{\rm in}$) in the spectral fitting of four models.
The data during the strong and weak flare periods are represented in orange and green color, respectively.
The solid red line corresponds to the equation $F_{\rm D}=c\cdot T_{\rm in}^{4}$, where $c$ is the best fit coefficient for the black data points. The upper and lower dashed lines represent the solid line multiplied by factors of 2 and 0.5, respectively.
\label{fig:DvsT}}
\end{figure}

The relationship between the disk bolometric flux ($F_{\rm D}$) and the inner radius of the multicolor disk ($R_{\rm in}$) is plotted in Figure \ref{fig:corrin}. Although the evolution trend of $R_{\rm in}$ calculated with model M2 significantly differs from the other three models during the weak flare period due to variations in $f_{\rm cov}$ in model M2, the four models show a positive correlation between $R_{\rm in}$ and $F_{\rm D}$ during the flare period.
Figure \ref{fig:DvsT} shows the relationship between $F_{\rm D}$ and $T_{\rm in}$ in the four models. During the weak flare period, M2 exhibits a tight relationship of $F_{\rm D}{\sim}T_{\rm in}^4$, indicating a relatively small range of variations in $R_{\rm in}$; while the other three models show significant dispersion, suggesting continuous evolution of $R_{\rm in}$. At the peak of the flare, all models demonstrate a rapid increase in $R_{\rm in}$ over a short period of time.

Taking all the above information into account, it can be seen that M1, M3, and M4 yield similar results, but differ from the results of M2. As M1 can be considered a mathematical simplification of the non-thermal radiation in M4, and M3 lacks a necessary non-thermal component compared to M4, we will mainly focus on comparing the results of models M2 and M4 in the following. 
For model M2, there are clear differences in the correlations between $R_{\rm in}$ and other parameters during strong and weak flare periods, while the correlations of other parameter pairs are similar during both periods, with only differences in the parameter values. 
This implies that the accretion disk states could change between the strong and weak flare periods.
For model M4, the correlation of each parameter pair is similar during both strong and weak flare periods, with only differences in the parameter values. This implies that there is no difference in the accretion disk state between strong and weak flare periods, and the flare is simply a result of rapid and significant changes in accretion disk parameters.
During the weak flare periods, $R_{\rm in}$ in M2 becomes very large and inversely correlated with $T_{\rm in}$, which is different from the decrease in $R_{\rm in}$ in M4. However, the results of M2 and M4 are consistent with each other during flare period. It can be concluded that flares originate from an increase in disk radiation, accompanied by an increase in accretion disk temperature and radius, as well as rapid changes in the radiation layer.

\subsection{Inner radius of the accretion disk}
\label{subsec:rin}

The change in $R_{\rm in}$ could be real, as instabilities in some accretion processes can lead to collapses in the inner regions of the accretion disk. However, it could also be apparent, merely reflecting changes in the hardening factor $f_{\rm col}$ during the evolution of the accretion disk. $f_{\rm col}$ is closely related to the state of the accretion disk, depending not only on the accretion process but also on the influence of disk-corona interactions \citep{2022ApJ...932...66R}.

Through the spectral analysis of these models, it can be seen that the flares are caused by a rapid increase in disk bolometric flux, which can be explained by various instabilities in mass transport \citep{1991ApJ...376..214B,2003PASJ...55L..69N,2013ApJ...777...11S}.
The thermal-viscous disk instability model previously used to understand dwarf nova outbursts has been extended to low-mass binary systems.
This instability is triggered by strong X-ray irradiation, which is easily influenced by the geometric properties of the accretion disk.
The X-ray radiation from the central source can heat the disk, leading to thermal instability affecting the mass transport of the accretion flow \citep{2001NewAR..45..449L}. This can cause the variations in viscosity and accretion efficiency, further result in fluctuations and instabilities in the disk. 
MHD simulations provide a valuable tool for elucidating the mass transport rates and viscosity. In three-dimensional MHD simulations, the magnetorotational instability generates turbulent motions that redistribute angular momentum and drive accretion onto the central black hole. The angular momentum transport rate is proportional to magnetic stress \citep{1996ApJ...463..656S,1998ApJ...501L.189A}. The mass transport rate in the accretion flow is directly related to the efficiency of angular momentum transport, which is determined by the viscosity coefficient $\alpha$. In Shakura-Sunyaev disk, $\alpha$ varies with radius in the weakly magnetized turbulent region outside the ISCO \citep{1973A&A....24..337S,2013MNRAS.428.2255P}. This variation is influenced not only by the turbulence but also by the geometric characteristics of the disk and corona. When a highly magnetized corona is located above a weakly magnetized accretion disk, their interaction may disrupt the inner region of the accretion disk \citep{2000ApJ...534..398M}. 
Therefore, it is reasonable for the accretion disk to collapse outward from the ISCO and lead to a rapid increase in $R_{\rm in}$ as Swift J1727.8--1613 is in the VHS with a high mass accretion rate.

The apparent increase in $R_{\rm in}$ caused by a decrease in $f_{\rm col}$ is another reasonable explanation, meaning that the change in $R_{\rm in}$ may not necessarily be physically real.
In the process of estimating $R_{\rm in}$ in Section 3, $f_{\rm col}$ was chosen as the typical value of 1.7 \citep{1995ApJ...445..780S}. However, many studies indicate that the hardening factor $f_{\rm col}$ varies in a range of 1.6--3 during the state evolution, related to parameters such as disk fraction and accretion rate. 
The spectral analysis with models M1--M4 indicated an overall gradual decrease in total flux (disk + non-thermal), which can be roughly considered as a decrease in the mass transfer rate. 
\cite{2008ApJ...683..389D} show that $f_{\rm col}$ is positively correlated with the accretion rate in the disk-dominated state.
By extending this conclusion to the observations presented here, i.e., $f_{\rm col}$ is positively correlated with the mass transfer rate in the VHS, it can be inferred that the hardening factor $f_{\rm col}$ continues to decrease.
The results of the M4k model clearly support this inference with an overall downward trend of $f_{\rm col}$ in the VHS. It is noteworthy that the hardening factor exhibits an overall anti-correlation with the disk accretion rate, which is in contrast to the positive correlation in the disk-dominated state \citep{2008ApJ...683..389D}. 
This indicates that the jets and the accretion disk are not completely independent in the VHS. They must have a direct physical relationship or interaction; otherwise, $f_{\rm col}$ and $\dot{M}$ would be expected to exhibit a positive correlation as shown by \cite{2008ApJ...683..389D}. 
\cite{2022ApJ...932...66R} have also pointed out that the PL component can make a non-negligible contribution to the hardening factor in the LHS and HIMS.
If the real $R_{\rm in}$ remains constant, then the inferred $R_{\rm in}$ under the diskbb model with a fixed $f_{\rm col}=1.7$ is expected to display an apparent upward trend, which is consistent with the overall trend of $R_{\rm in}$ shown in the appendix.
Actually, $R_{\rm in}$ may be stable at the ISCO in the VHS. \cite{2011MNRAS.411..337D} found that $f_{\rm col}$ is inversely related to the disk fraction in the disk-dominated state. This rule is used to analyze the data during strong and weak flare periods, although our data does not indicate disk domination. During strong flare periods, the data has a higher accretion disk proportion compared to weak flare periods, and $f_{\rm col}$ is relatively lower. If $R_{\rm in}$ remains constant, the apparent $R_{\rm in}$ during strong flare periods will also be greater than during weak flare periods. This phenomenon is most significant at the peak of the flare. This is consistent with our results during the results of model M2 in the strong flare period and model M4 in the entire period, but is in conflict with the result of model M2 in the weak flare periods. Therefore, it can be inferred that within the framework of M2 model, the radius does indeed increase during the weak flare periods.
In conclusion, the overall gradual increase in $R_{\rm in}$ can be the result of the long-term evolution of hardening factors $f_{\rm col}$, and the rapid rise in $R_{\rm in}$ around flare peaks can be caused by short-term variations in $f_{\rm col}$. Models M2 and M4 have a consistent understanding of the variation in $R_{\rm in}$ during the flare periods, where $R_{\rm in}$ remains at the ISCO, and the apparent changes are due to variations in $f_{\rm col}$. They have different interpretations of $R_{\rm in}$ during the weak flare periods. Model M2 suggests that the increase in $R_{\rm in}$ is real and underestimated due to the use of small $f_{\rm col}$; while in model M4, $R_{\rm in}$ can still be considered constant, with its apparent changes explained by variations in $f_{\rm col}$. 

\subsection{Mechanism of the Non-Thermal Radiation}
\label{subsec:non-th}

A bump in the spectrum around 10 to 20~keV is fitted by the Comptonization model in M2 model, while a hard tail in the higher energy band above 100~keV is fitted with a PL component. A natural idea is that the PL component without a cut-off in model M2 originates from the inverse Compton process of non-thermal distribution of electrons. 
Therefore, simple thermal or non-thermal electron Comptonization processes cannot explain the observed spectra, and more complex electron distribution scenarios are required.
The similar scenario in XTE J1550--564 is described by the Comptonization process with a hybrid distribution of thermal and non-thermal electrons \citep{2003MNRAS.342.1083G}.
In our spectral analysis, the thermal electrons follow a Maxwellian distribution with a temperature around 15 to 30 keV, contributing to Comptonization process in the lower energy band; while the non-thermal electrons with higher energies undergo Comptonization process in the higher energy band to generate a PL continuum extending up to several hundred keV in hard X-rays.
In model M2, the Comptonization process by the thermal and non-thermal electron distributions are considered to be independent of each other.
The fitting results of model M4 and the absence of iron emission indicate a weak reflection component, implying that the intense PL component does not illuminate the accretion disk, which also imposes strong constraints on the origin of the PL component.
Additionally, the polarization observations \citep{2023ApJ...958L..16V,2024ApJ...966L..35S} suggest that these electrons involved in the Compton process should be located in the inner region of the accretion disk and have relatively low height.

\begin{figure}[ht!]
\centering
\includegraphics[scale=0.5]{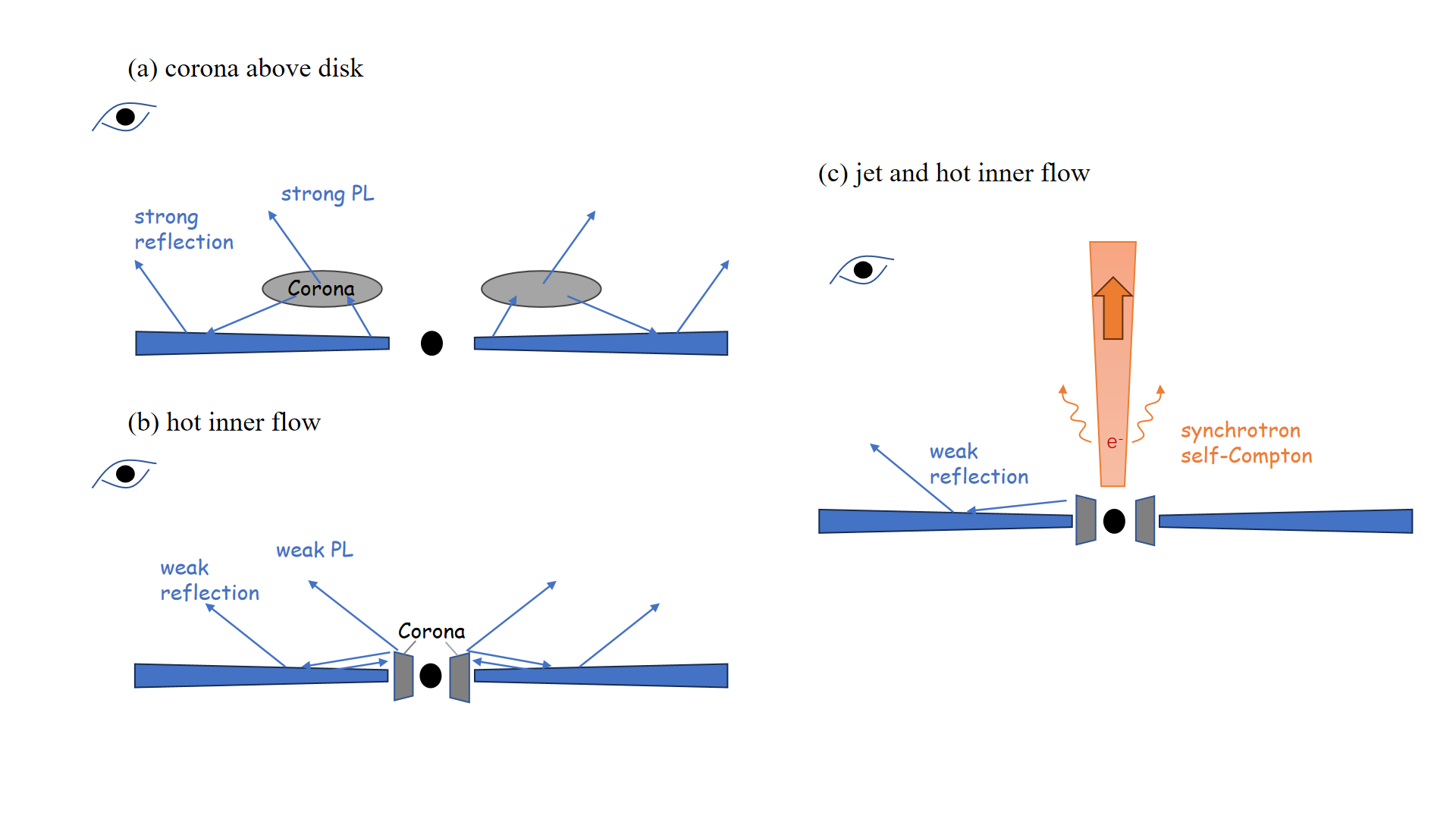}
\caption{Schematic diagrams of the geometries of the disk and corona. Panel (a): A corona above the disk can contribute to a strong PL, and irradiate the disk resulting a strong reflection. Panel (b): A weak hot inner flow located in the inner disk region can contribute to a weak PL. A small fraction of hard photons irradiate the disk, resulting in a weak reflection. Panel (c): The disk with a weak hot inner flow can contribute to a weak reflection and PL, and the synchrotron self-Compton in a relativistic jet can produce a strong PL.
\label{fig:pysmodel}}
\end{figure}

\begin{figure}[ht!]
\centering
\includegraphics[scale=0.4]{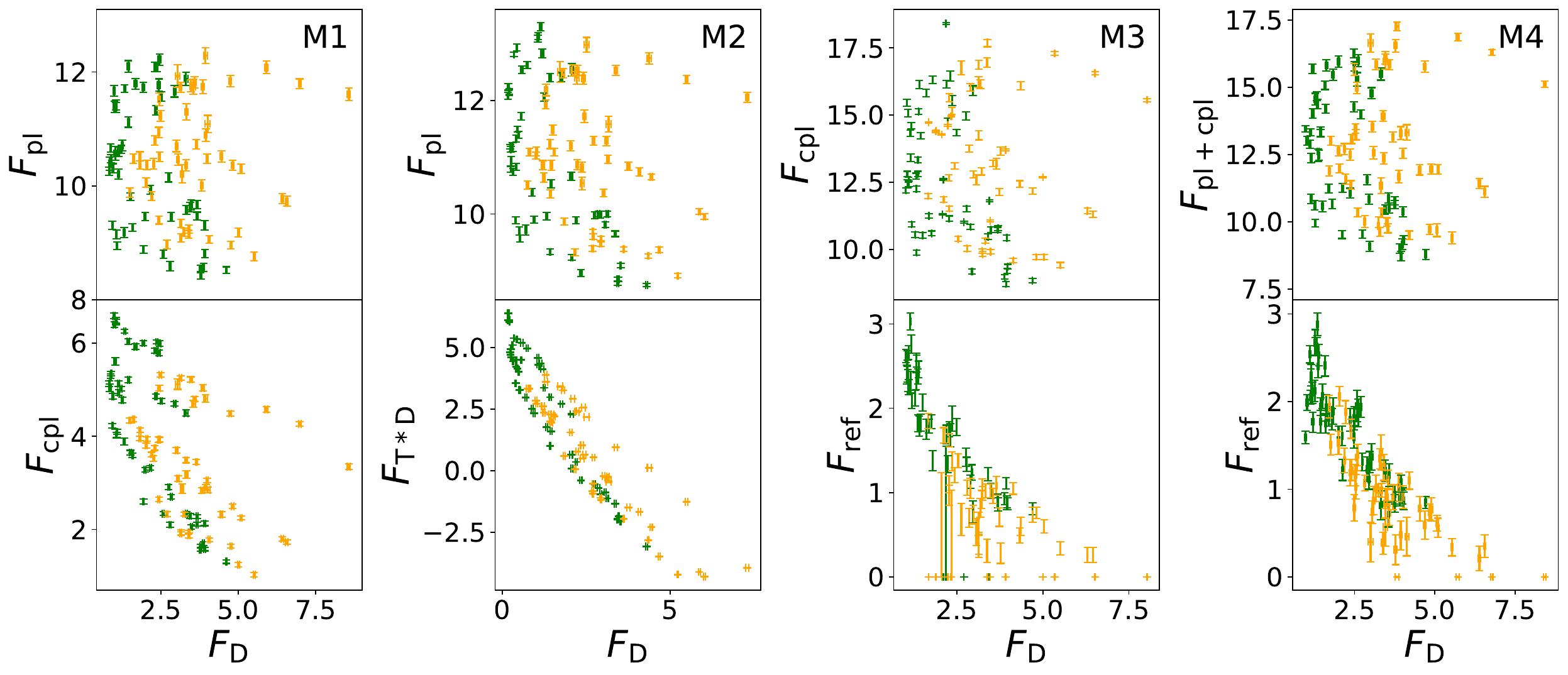}
\caption{
Top: The relationship between the disk bolometric flux and the flux of the first non-thermal component in the four models. Bottom: same as the top panels but for the second non-thermal component. $F_{\rm PL}$, $F_{\rm CPL}$, $F_{\rm ref}$, $F_{\rm T*D}$ and $F_{\rm D}$ are the flux of power-law, cutoff power-law, reflection component, Comptonized and non-Comptonized multi-color disk, respectively, with units all in $\rm 10^{-8}~erg~cm^{-2}~s^{-1}$. The plotting colors follows the conventions of Figure~\ref{fig:lcurve}.
\label{fig:parrelation}}
\end{figure}

Several potential disk-corona geometries are discussed as follows, illustrated by the cartoons in Figure~\ref{fig:pysmodel}. As shown in panel (a) of Figure~\ref{fig:pysmodel}, the corona is assumed to be located on the accretion disk and contributes to the strong PL component. Therefore, the hard photons from the corona can easily irradiate the disk and lead to a strong reflection feature that conflicts with the weak reflection observed.
In the scenario shown in panel (b) of Figure~\ref{fig:pysmodel}, where a small geometric scale hot inner flow located between disk and black hole, only a small fraction of disk photons undergo Comptonization in the corona and are scattered upwards to high energies. This can results in weak coronal radiation and a weak reflection component, which can explain the CPL component and weak reflection component in model M4, as well as the Comptonized disk with no reflection component in model M2. However, this cannot explain the strong PL component that dominates the spectrum, indicating that the strong PL component has another origin.
Relativistic jets are typically regarded as being generated by the acceleration of matter around a black hole or accretion disk under the influence of a strong magnetic field \citep{1977MNRAS.179..433B,1982MNRAS.199..883B},
and the synchrotron radiation from the jet could be a more reasonable source of the PL component \citep{2005ApJ...635.1203M}.
Due to the high-speed outward bulk motion of electrons within the jet, the disk is unable to provide seed photons for the inverse Compton process. The synchrotron radiation of relativistic electrons in the magnetic field within the jet can be a plausible source of seed photons \citep{2013MNRAS.430.3196V}. These synchrotron photons then undergo inverse Compton scattering with relativistic electrons, known as the synchrotron self-Compton process, generating a strong PL component. Since the direction of the jet radiation being outward from the accretion disk, the strong PL component cannot illuminate the accretion disk and cannot contribute to the reflection component \citep{2021NatCo..12.1025Y}. 
Additionally, the radio VLBA observations have confirmed the presence of the jet \citep{2024ApJ...971L...9W}. In the spectral analysis conducted with \textit{Insight}-HXMT, the spectrum of Swift J2727.8--1613 can be simply divided into two main components, i.e., the disk-corona interaction system and the jet base. The polarization angle of the radiation produced by the disk-corona interaction system in the soft X-ray energy range is approximately parallel to the direction of the jet \citep{2024ApJ...968...76I}, and the synchrotron self-Compton process at the jet base will produce unpolarized soft X-ray radiation. Therefore, the polarization features in soft X-ray mainly come from the disk-corona interaction system, which are consistent with 
our physical scenario (Figure~\ref{fig:pysmodel}).
As shown in Figure \ref{fig:parrelation}, there is no significant correlation between the PL component and the disk component, while a clear anti-correlation between the disk component flux and the scattered or reflected component flux is discovered. The correlations between components are consistent with the accretion disk with a hot inner flow and jet model shown in Figure \ref{fig:pysmodel}. The significant high-energy hard component does not originate from the corona above the disk or the hot inner flow within the inner disk region, but rather from synchrotron self-Compton radiation in the relativistic jet. 
There is no fundamental difference in the physical scenarios between models M2 and M4, except in the description of the non-thermal component generated by the hot inner flow. This has led to different interpretations of some data during weak flare periods, such as the evolution of $R_{\rm in}$ shown in appendix and its relationship with other parameters (Figure~\ref{fig:corr}--\ref{fig:corrin}).

\subsection{Comparison with GRO J1655--40 and XTE J1550--564}

Many XRBs have been historically observed in the VHS or the so-called SPL state. During the outbursts of GRO J1655--40 in 1996--1997 and XTE J1550--564 in 1998--1999, both sources underwent transitions from the VHS to the HSS in RXTE spectral observations \citep{1999ApJ...520..776S,2000ApJ...544..993S}.
The spectra in the VHS are dominated by a strong PL component, while the spectra in the HSS are dominated by the disk component. In the VHS, some flares were observed, accompanied by a sudden increase in the PL component and a sharp decrease in  $R_{\rm in}$.
During the flare periods of GRO J1655--40 and XTE J1550--564, the observed decrease in $R_{\rm in}$ with the increase of the PL component can be understood as an increase in $f_{\rm col}$. The positive correlation between the accretion rate and hardening factor is obvious, and it can explain the overall increase in $R_{\rm in}$ observed in Swift J1727.8--1613, where the PL component shows an overall slow decreasing trend accompanied by a long-term decrease in $f_{\rm col}$. Unlike GRO J1655--40 and XTE J1550--564, the flares in Swift J1727.8--1613 originate from an increase in the disk bolometric flux, and the apparent increase in $R_{\rm in}$ can be attributed to a lower $f_{\rm col}$ during the strong flare periods compared to weak flare periods.
Based on the spectral characteristics of GRO J1655--40 and XTE J1550--564, a truncated accretion disk with a Comptonized corona is generally considered to exist around the black hole \citep{1997ApJ...489..234H,2006csxs.book..157M}.
The observation of XTE J1550--564 in the VHS suggests that the inner edge of the accretion disk is truncated at a radius greater than the ISCO, possibly due to a strong Comptonized corona completely covering the accretion disk.
The complex geometry of the intense Comptonized corona leads to changes in the colour temperature correction factor, influencing the calculation of the effective radius \citep{2004MNRAS.353..980K}. 
Furthermore, a hard tail in the high-energy range was identified, which can be explained using a strong Comptonization model incorporating a hybrid distribution of thermal and non-thermal electrons \citep{2003MNRAS.342.1083G}. 
In our spectral analysis, a comparable high-energy hard tail was also detected.
However, the absence of correlation between the PL and the disk component during the VHS of Swift J1727.8--1613 casts doubt in explaining this hard tail by the hybrid plasma model.
The independent PL component, which dominates the spectrum during the VHS, is thus suggested to originate from the synchrotron self-Compton in a jet structure and contributes very little to the reflection component, as shown in panel (c) of Figure~\ref{fig:pysmodel}.

\begin{figure}[ht!]
\centering
\includegraphics[scale=0.4]{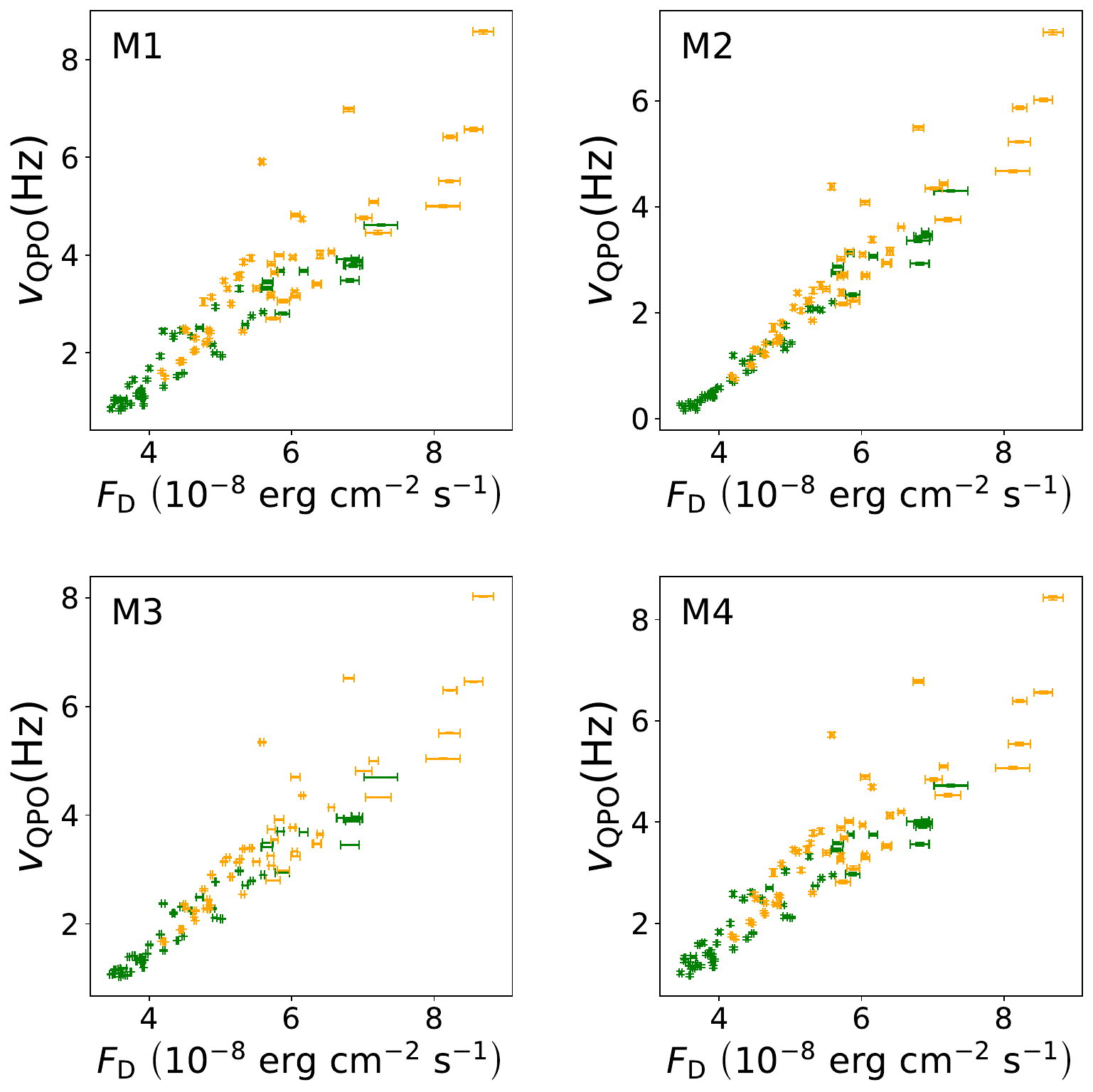}
\caption{The relationship between the central frequency of low-frequency QPO  ($\nu_{\rm QPO}$) and the disk bolometric flux ($F_{\rm D}$) in the spectral fitting of four models.
\label{fig:qpo}}
\end{figure}

During the 2023 outburst of Swift J1727.8--1613, a series of low-frequency QPOs have been observed \citep{2024MNRAS.529.4624Y,2024ApJ...968..106Z,2024MNRAS.531.1149N}, all of which are typical Type-C QPOs during the flare period. 
A significant relation between the central frequency of LFQPO and $T_{\rm in}$ is exhibited in Figure \ref{fig:qpo}. 
The similar relation has been observed in other BHXRBs, e.g., GRO J1655--40 and XTE J1550--564.
GRO J1655--50 exhibits variable QPOs with central frequencies of 14--22~Hz, and which are exhibited in XTE J1550--564 with central frequencies of 0.08--18~Hz. 
\cite{2000ApJ...531..537S} researched a strong positive correlation between the QPO frequency and disk bolometric flux in the above QPO, indicating the QPO frequency increases as the mass accretion rate of disk increases. 
A popular explanation for QPOs in the past was that they originated from the relativistic (Lense-Thirring) precession of an inner accretion flow \citep{2019NewAR..8501524I}, however, this view has been challenged in recent years.
\cite{2022MNRAS.511..255N} observed a significant modulation in the reflection fraction during the observations with NICER and NuSTAR, supporting the geometric origin of the QPO. Nevertheless, the inferred inner disk radius is too small to be explained by Lense-Thirring precession for the observed QPO frequency.
Some theoretical analyses also have raised doubts about whether the inner accretion flow can exhibit Lense-Thirring precession \citep{2021ApJ...906..106M}.
Jet precession is also a possible source of QPOs, which successfully explains the soft phase lag observed in MAXI~1820+070 \citep{2021NatAs...5...94M}.

\section{Summary} \label{sec:summary}
Swift J1727.8--1613 underwent a strong outburst in 2023.
After a brief LHS at the beginning, it transitioned into the HIMS, which can be seen from the HID and the temporal characteristics. As a series of flares are displayed on the light curve in soft X-ray, it entered the VHS. Through the observation result with IXPE and MAXI data \citep{2024ApJ...966L..35S,2024ATel16541....1P}, it entered the soft state, and then transitioned back to the LHS.
\textit{Insight}-HXMT conducted high-cadence monitoring of Swift J1727.8--1613 before it entered the soft state, and observed a series of significant flares in the soft X-ray energy range. We focus on the spectral evolution during the flare period.

Among the models investigated in Section~\ref{sec:result}, the Comptonized disk of model M2 and the disk reflection of model M4 both have clear physical pictures and provide statistically acceptable fitting goodness. Additionally, model M4k, which includes the relativistic accretion disk model kerrbb, was also examined as an important supplement to model M4.
From these spectral fittings with different models, we have obtained the following consistent and definitive conclusions.

In the state transition from the HIMS to the VHS (first flare: MJD 60197--60204):
\begin{enumerate}
\item The inner edge of the accretion disk has a low temperature $T_{\rm in}\sim0.5$~keV, but a high temperature $T_{\rm in}\sim1$~keV is observed near the peak of the flare, which is close to that in the VHS.
\item The inner radius of the accretion disk decreases as the flare increases until it reaches the ISCO when the flare is at its peak, and then increases as the flare decreases. It approaches the ISCO after the source enters the VHS. 
\item The disk component contributes a very low proportion of the total flux, with the non-thermal component dominating the total flux. Flares are mainly contributed by the power-law non-thermal component with a photon index of 2.0--2.6, which falls between the values of the LHS and VHS.
\end{enumerate}

In the VHS (MJD 60204--60220):
\begin{enumerate}
    \item The spectrum during the VHS includes a disk component and at least two non-thermal components, with the non-thermal components dominating the total flux. 
    The disk component has the lowest proportion of the total flux, but the proportion of the accretion disk increases from about 10\% to 30\% as the radius of the accretion disk expands when the flare intensity reaches its maximum. Overall, the non-thermal component dominates the total flux in 2--200 keV.
    \item The PL component accounts for over half of the total flux, and it has a very steep spectral shape with a large spectral index $\alpha_{\rm PL}\sim2.6$. It maintains a stable spectral shape during the VHS, with only a slow decrease in flux. However, the very weak reflection and lack of correlation with the disk component contradicts the conventional explanation of the PL component as the inverse Comptonization of disk photons.
    \item The accretion disk with a small scale hot inner flow and jet scenario is the most reasonable explanation for the \textit{Insight}-HXMT data. The strong PL component originates from the synchrotron self-Compton process in the relativistic jet, while the other weak non-thermal component is contributed by the hot inner flow of the accretion disk.
    \item The flares originate from rapid and intense increase in accretion disk emission, along with an increase in $R_{\rm in}$ and $T_{\rm in}$. The variation in $R_{\rm in}$ can be real, as the scenario has been confirmed by both theoretical and simulation studies. However, it may not necessarily be physical, and it could be caused by changes in $f_{\rm col}$ with the variation of disk fraction \citep{2011MNRAS.411..337D}. 
\end{enumerate}


\begin{acknowledgments}
This work made use of the data from the \textit{Insight}-HXMT mission, a project funded by the China National Space Administration (CNSA) and the Chinese Academy of Sciences (CAS). The authors are thankful for the support from the National Natural Science Foundation of China under grant Nos. 12333007, U2031205, 12403053 and the National Key R\&D Program of China (grant No. 2021YFA0718500). This work was partially supported by the International Partnership Program of the Chinese Academy of Sciences (grant No. 113111KYSB20190020). 
\end{acknowledgments}

\appendix

\renewcommand{\thefigure}{{A}\arabic{figure}}
\setcounter{figure}{0}

\begin{figure}[ht!]
\centering
\subfigure{\includegraphics[scale=0.22]{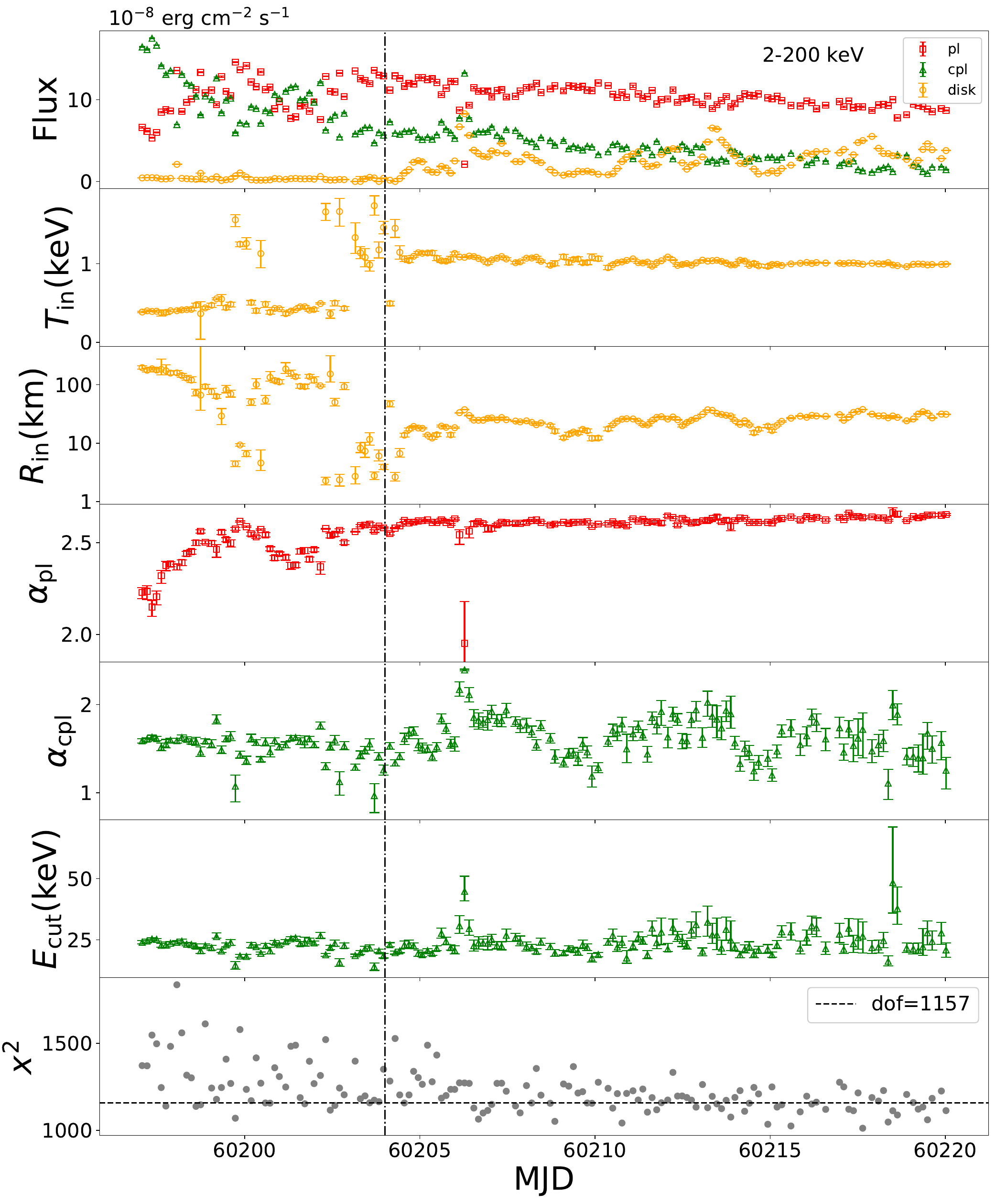}}
\hspace{.06in}
\subfigure{\includegraphics[scale=0.22]{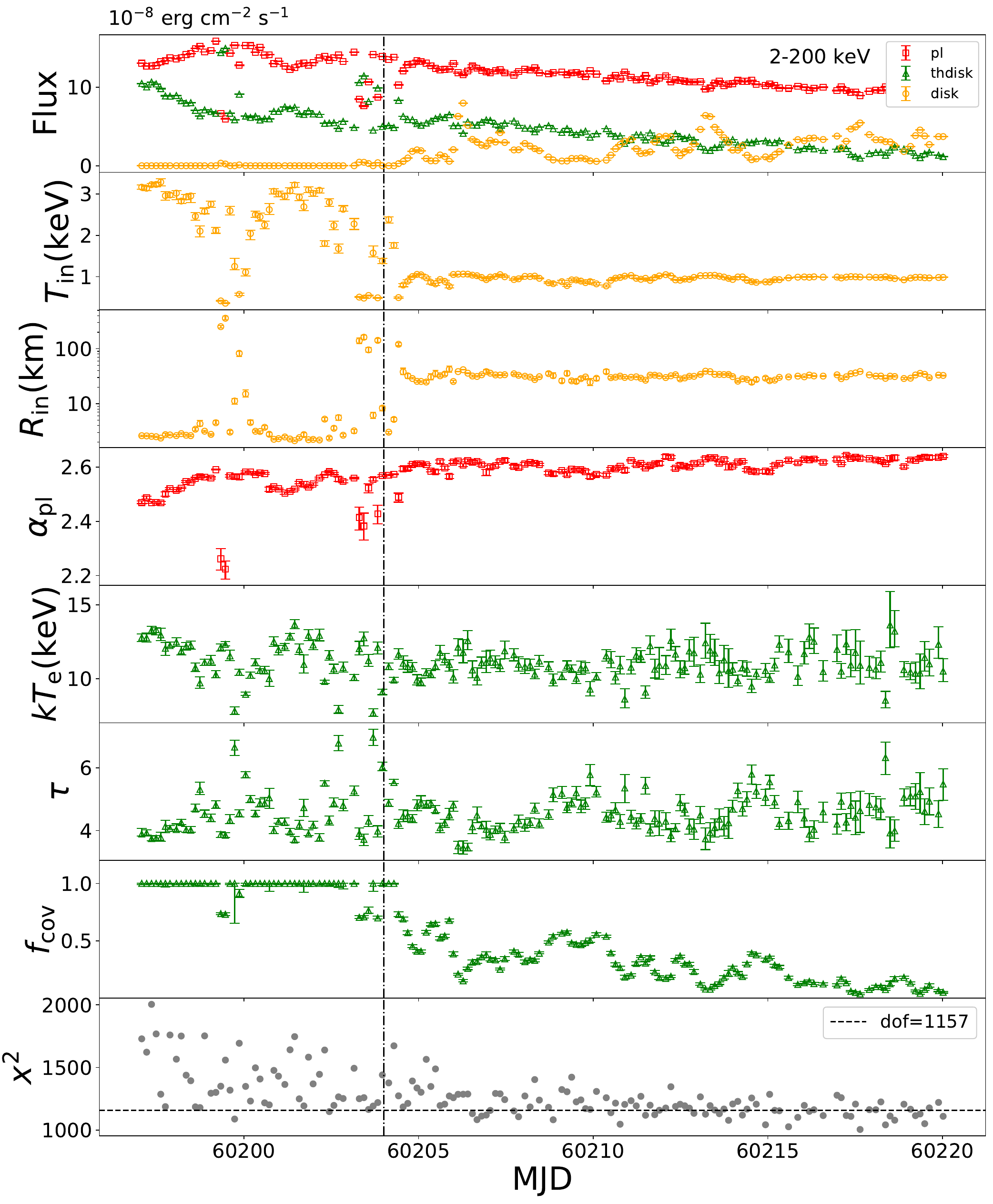}}
\caption{
The spectral parameters and fitting goodness obtained from the spectral fitting with models M1 (left) and M2 (right).
The vertical dotted line indicates the time when the source transitions to the VHS.
The orange, red and green colors represent the parameters related to the disk, power-law spectrum, and other non-thermal components, respectively.
\label{fig:prefit_m1-2}}
\end{figure}

\begin{figure}[ht!]
\centering
\subfigure{\includegraphics[scale=0.22]{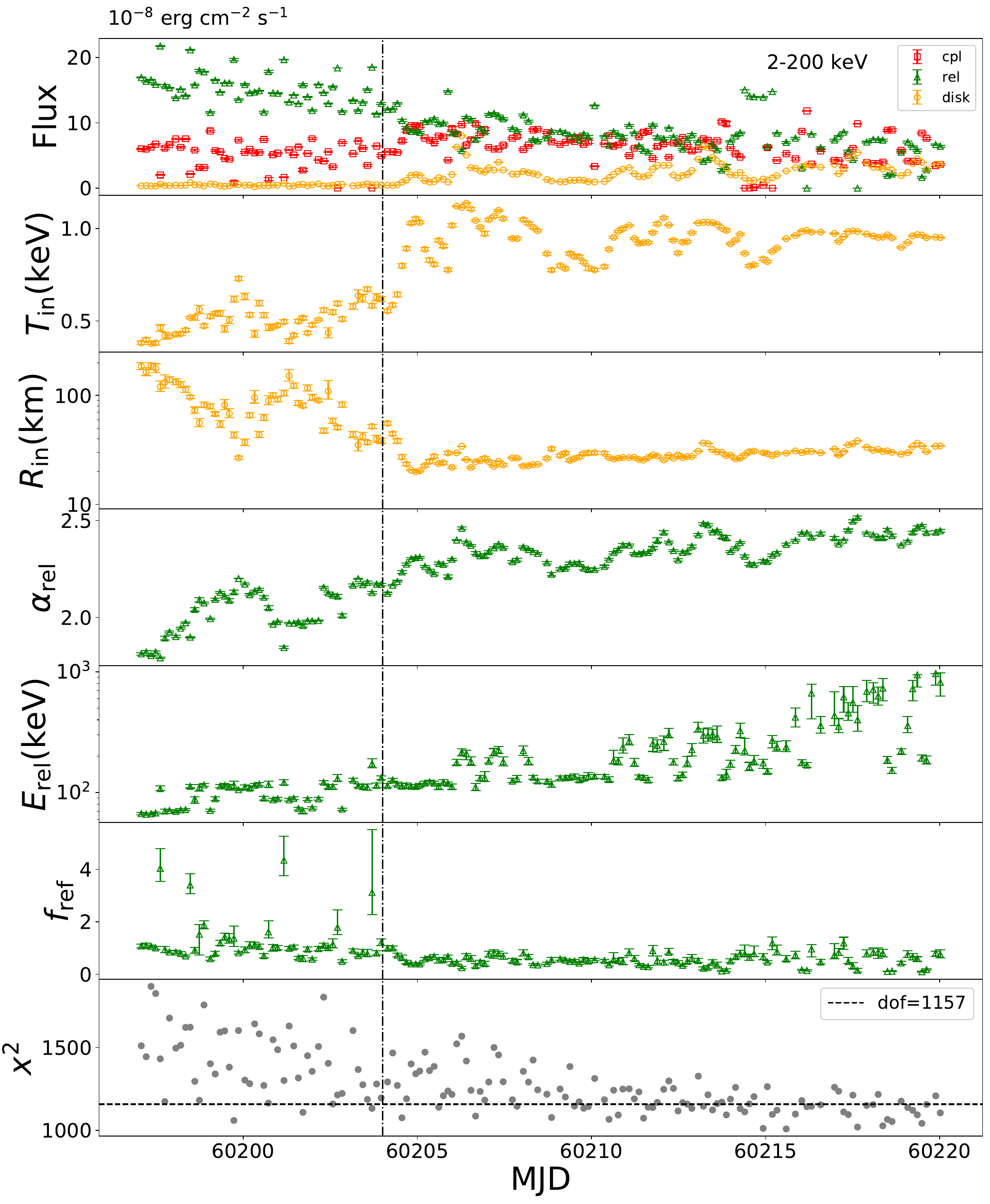}}
\hspace{.06in}
\subfigure{\includegraphics[scale=0.22]{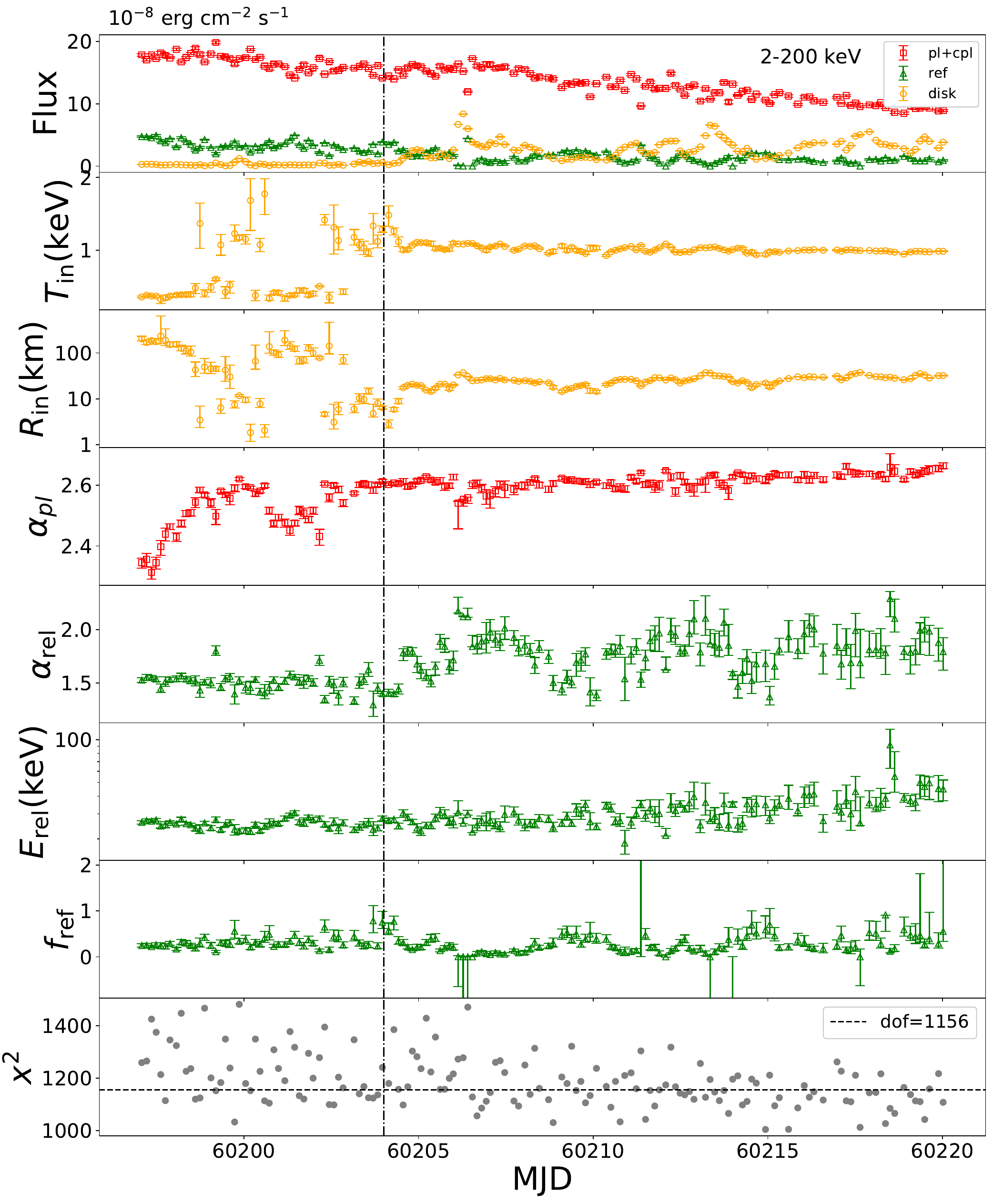}}
\caption{
Same as Figure~\ref{fig:prefit_m1-2} but with models M3 (left) and M4 (right).
\label{fig:prefit_m3-4}}
\end{figure}

\begin{figure}[ht!]
\centering
\subfigure{\includegraphics[scale=0.22]{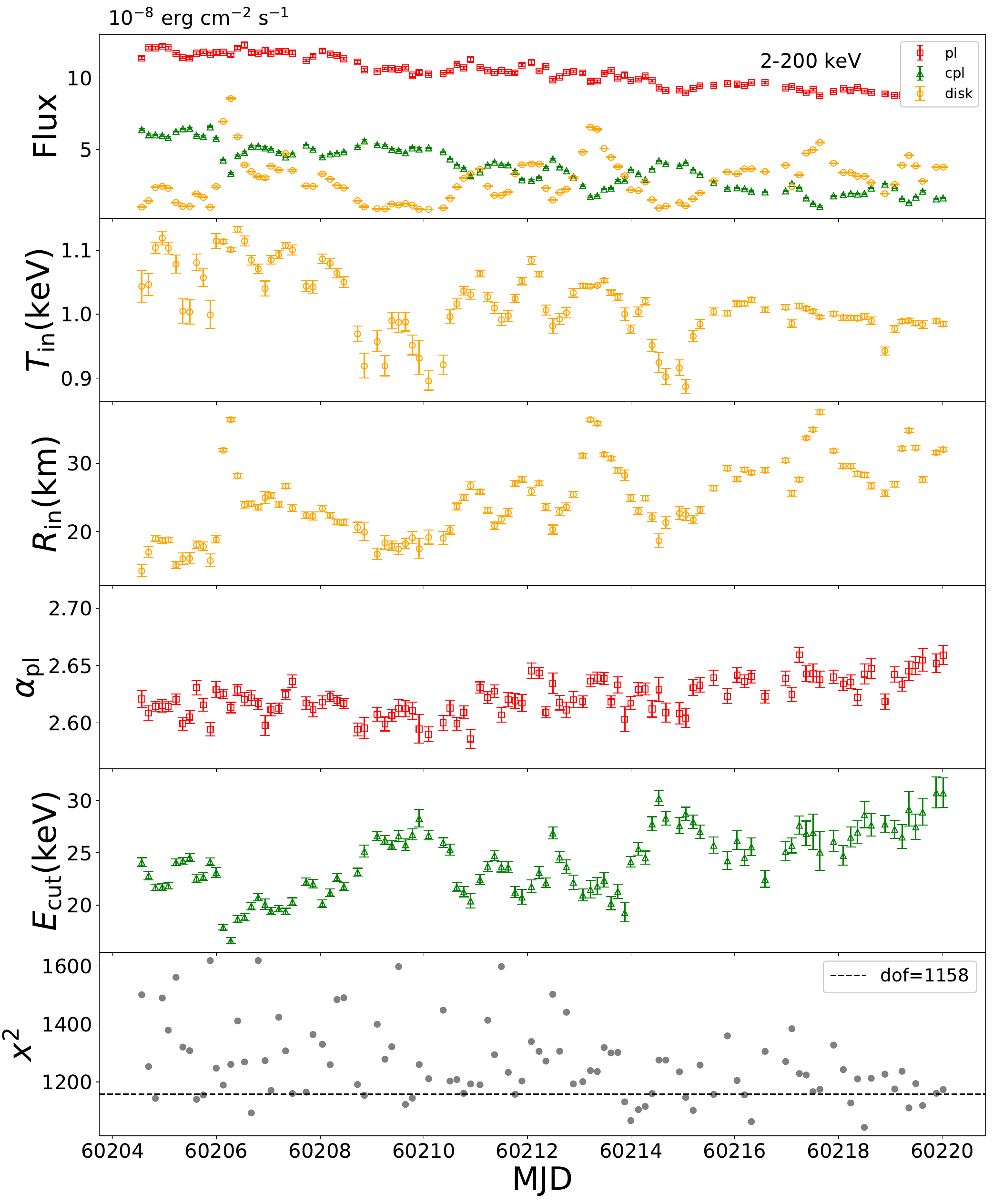}}
\hspace{.06in}
\subfigure{\includegraphics[scale=0.22]{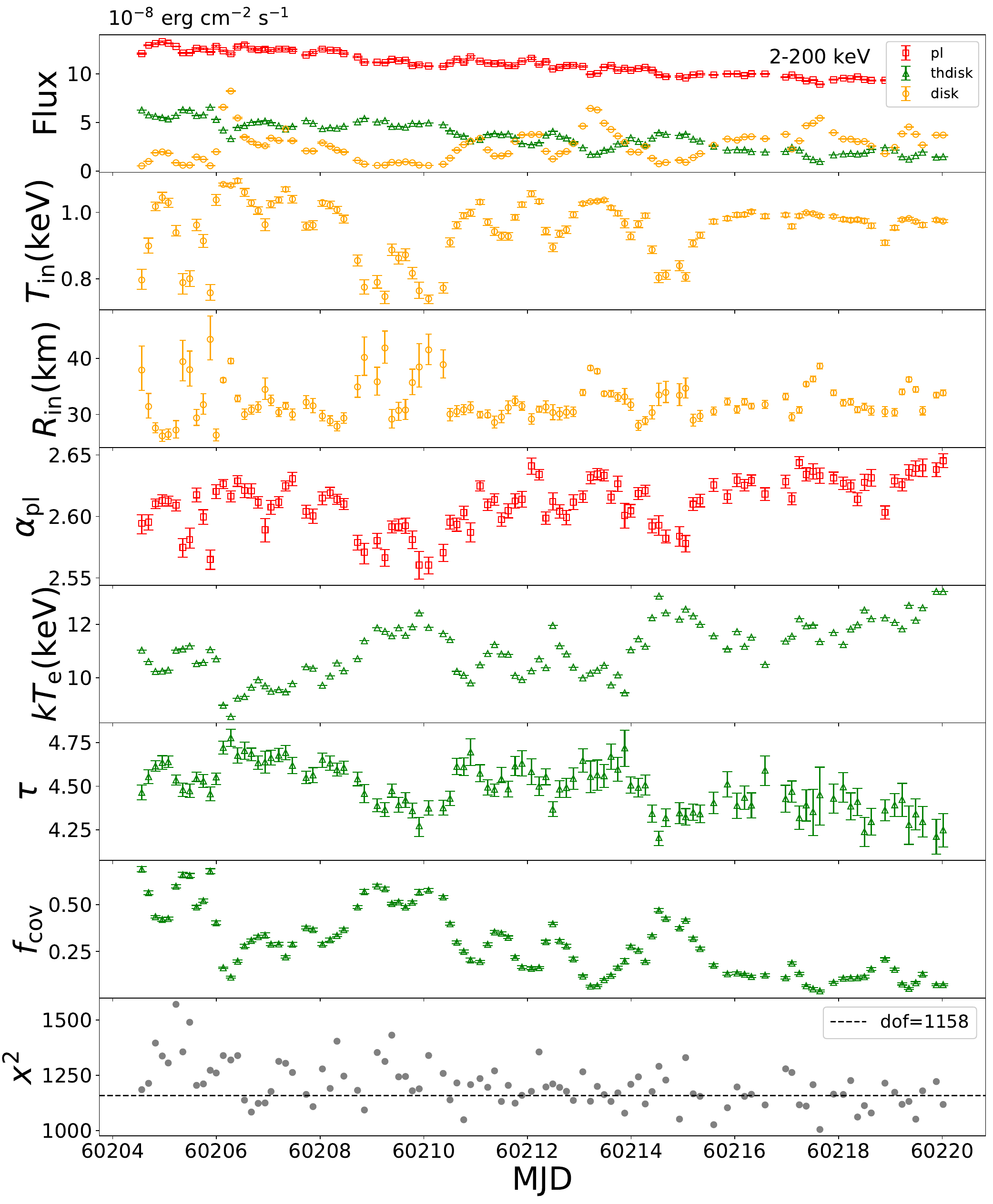}}
\caption{
The spectral parameters and fitting goodness obtained from the spectral fitting for the source in the VHS fitted with models M1 (left) and M2 (right). 
Some parameters are frozen to reduce degeneracy.
The plotting colors follows the conventions of Figure~\ref{fig:prefit_m1-2}.
\label{fig:m1-m2}}
\end{figure}

\begin{figure}[ht!]
\centering
\subfigure{\includegraphics[scale=0.22]{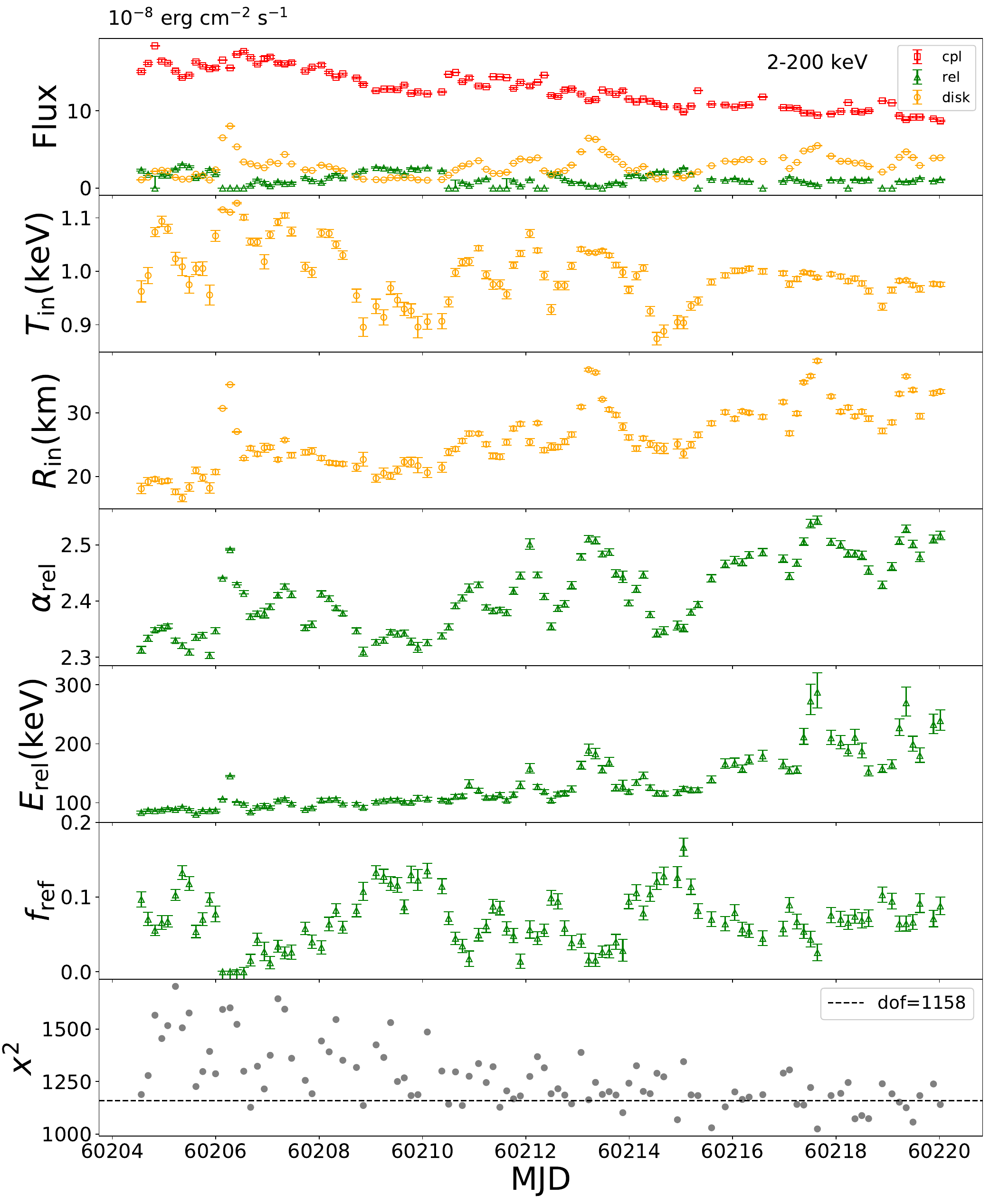}}
\hspace{.06in}
\subfigure{\includegraphics[scale=0.22]{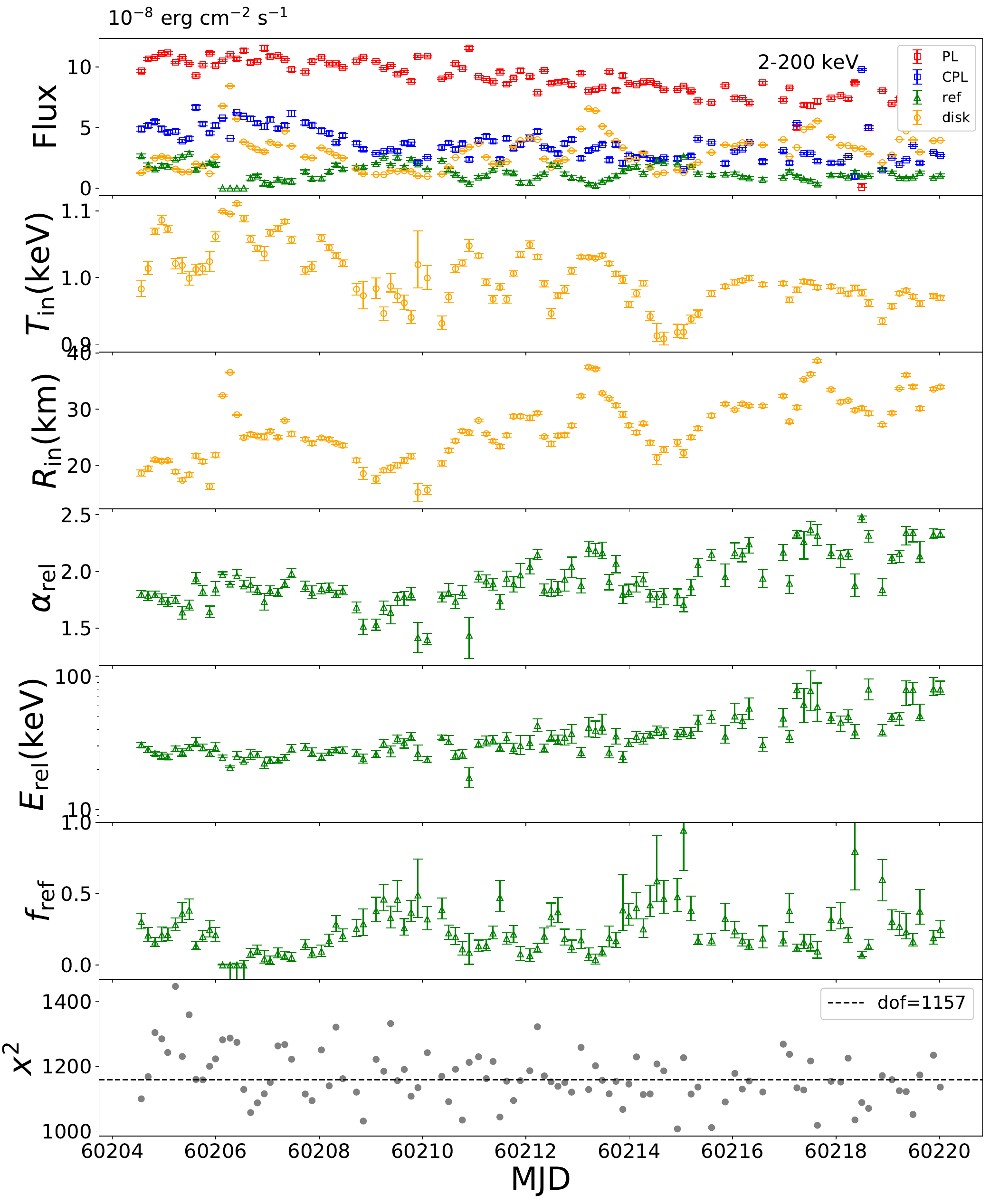}}
\caption{Same as Figure \ref{fig:m1-m2} but with models M3 (left) and M4 (right). The flux of CPL component $F_{\rm CPL}$ in the left panel (red color) and the right panel (blue color) both are obtained from the incident spectrum in the relxill model.
\label{fig:m3-m4}}
\end{figure}

\begin{figure}[ht!]
\centering
\subfigure{\includegraphics[scale=0.22]{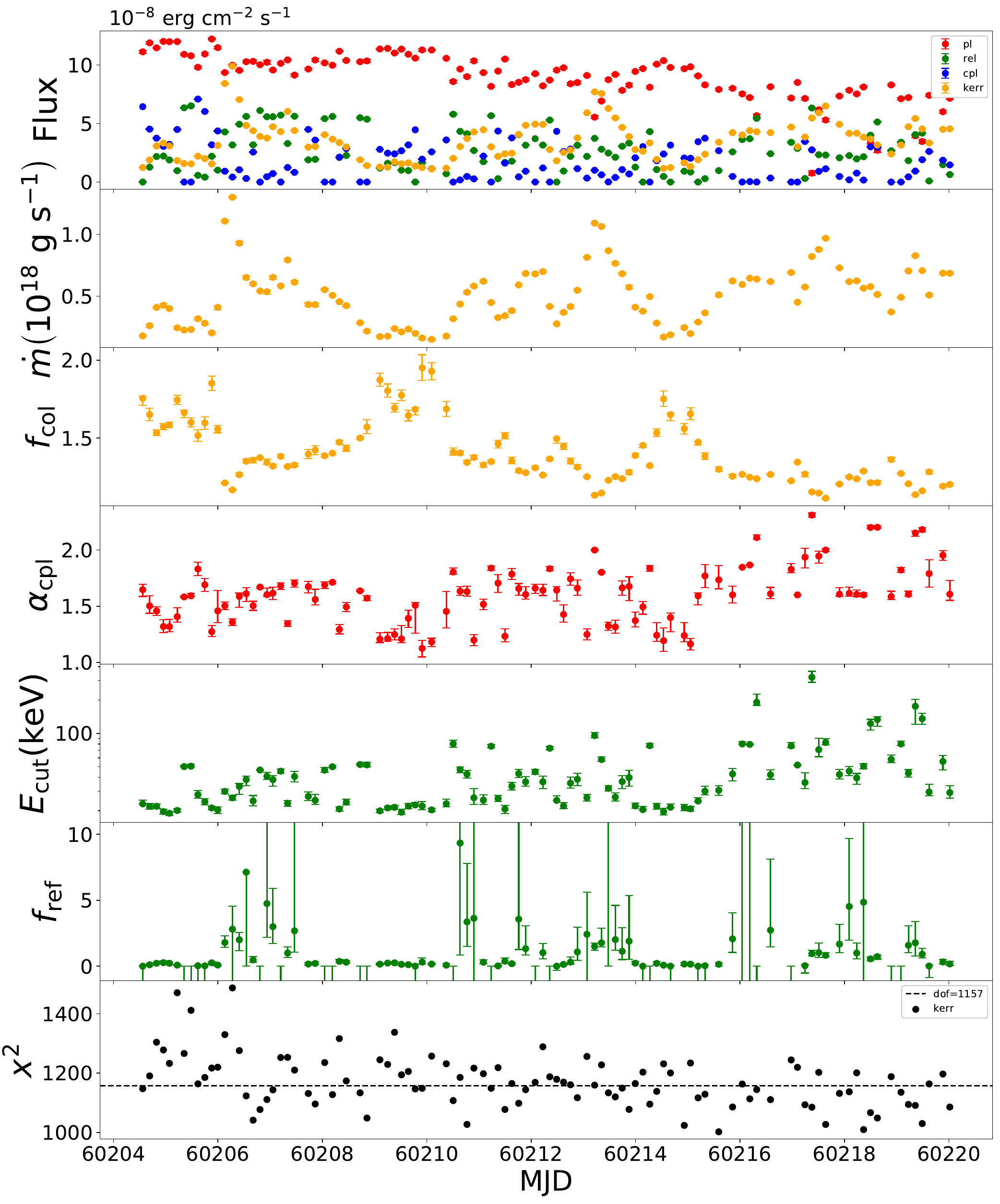}}
\caption{Same as Figure \ref{fig:m1-m2} but with model M4k.
\label{fig:m4k}}
\end{figure}

Figures \ref{fig:prefit_m1-2} and \ref{fig:prefit_m3-4} display the overall spectral evolution with four models during the period from MJD 60197 to MJD 60220, respectively. The spectral analysis results of models M1--M4 with most parameters free in fitting suggest the possible presence of parameter degeneracy. More detailed description and discussion are in Section \ref{sec:st}.
Therefore, in order to reduce the potential parameter degeneracy, we fixed some parameters when fitting the spectra during the VHS with four models. The results of the spectral fitting are shown in Figures \ref{fig:m1-m2} and \ref{fig:m3-m4}. In Section \ref{sec:vhs}, we described the specific parameter fixing conditions for each model and presented the results of the spectral analysis.
As a meaningful supplement, Figure A5 illustrates the parameter evolution of model M4k, which is derived from replacing the diskbb model in model M4 with the relativistic accretion disk model kerrbb.


\bibliography{sample631}{}
\bibliographystyle{aasjournal}



\end{CJK*}
\end{document}